\documentclass[journal=mmbox,manuscript=communication,layout=twocolumn]{achemso}
\usepackage[version=3]{mhchem} 
\usepackage{achemso,amsmath,relsize}

\def\D{\mathrm{d}}

\newcommand{\imun}{i\:}

\author{Doros N. Theodorou}
\email{doros@central.ntua.gr}
\affiliation[NTUA]{School of Chemical Engineering, National Technical University of Athens, 
9 Heroon Polytechniou Street, Zografou Campus, GR-15780 Athens, Greece}
\author{Georgios G. Vogiatzis}
\author{Georgios Kritikos}

\title{Self-Consistent-Field Study of Adsorption and Desorption Kinetics of Polyethylene Melts 
on Graphite and Comparison with Atomistic Simulations}

\makeatletter
\let\thetitle\@title
\let\theauthor\@author
\makeatother

\makeatletter
\renewcommand\section{\@startsection{section}{1}{\z@}%
                                  {-3.5ex \@plus -1ex \@minus -.2ex}%
                                  {2.3ex \@plus.2ex}%
                                  {\normalfont\small\bfseries}}                                  
\makeatother

\makeatletter
\renewcommand\subsection{\@startsection{subsection}{1}{\z@}%
                                  {-3.5ex \@plus -1ex \@minus -.2ex}%
                                  {2.3ex \@plus.2ex}%
                                  {\normalfont\small\small\bfseries}}
\makeatother

\begin{document}
\begin{abstract} 
\begin{figure} 
\begin{center}
  \includegraphics[clip,width=1.0\linewidth] {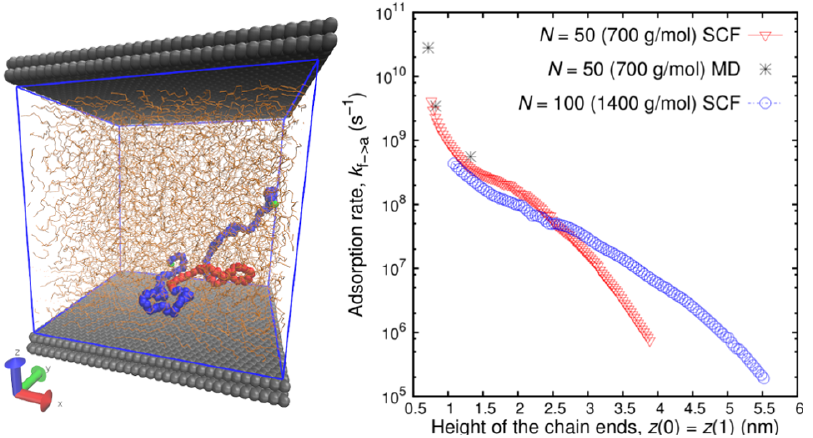}
\end{center}
\end{figure}

A method is formulated, based on combining self-consistent field theory with dynamically corrected
transition state theory, for estimating the rates of adsorption and desorption of end-constrained chains 
(e.g. by crosslinks or entanglements) from a polymer melt onto a solid substrate.  
This approach is tested on a polyethylene/graphite system, where the whole methodology is parametrized by atomistically 
detailed molecular simulations. For short-chain melts, which can still be 
addressed by molecular dynamics simulations with reasonable computational resources, the self-consistent 
field approach gives predictions of the adsorption and desorption rate constants which are gratifyingly
close to molecular dynamics estimates. 
\end{abstract} 


\section{Introduction}

Understanding the fascinating and complex dynamics of melts of large flexible polymer coils close to solid substrates
has been an ongoing challenge for many decades. 
From the point of view of molecular simulations, the spectrum of length and time scales associated with polymer melts 
of long chains poses a formidable challenge to studying the long-time dynamics.\cite{ChemEngSci_62_5697,MolPhys_102_147} 
The topological constraints arising from chain connectivity and uncrossability (entanglements) dominate 
intermediate and long-time relaxation\cite{Macromolecules_45_9475}
and transport phenomena when polymers become sufficiently long.
Atomistic molecular simulations of dense phases of soft matter prove to be difficult for many systems across 
length and time scales of practical interest. Even coarse-grained particle-based simulation methods may not be 
applicable due to the lack of faithful descriptions of polymer-polymer and polymer-surface interactions. 
While the study of polymer adsorption from solution onto a solid surface has a long 
history,\cite{JPhysChem_57_584,JChemPhys_43_2392} the adsorption from a melt has been less studied.

The thermophysical and rheological properties of polymer melts are of paramount importance in plastics processing.
Many thermoplastic products are formed by means of processes in which the molten polymer flow is limited by solid
surfaces (e.g. in extruders or in compression molding). 
The rheological behavior of a polymer melt may be well known, but unexpected phenomena such as sharkskin instability 
and melt fracture, which are intimately related to polymer-substrate interactions, 
occur at the interface of the melt and the mold or extruder.\cite{PolymEngSci_38_101,Polymer_Melt_Fracture}
Polymer chains can adsorb on surfaces and modify their effective interaction as well as resistance to their relative 
motion, a phenomenon with consequences in areas which range from tribology to biology. 
It has been theoretically proposed that, under certain conditions, a highly entangled polymeric fluid may not maintain 
the usual stick boundary condition at a solid wall.\cite{Langmuir_8_3033}
A significant and unambiguous violation of the law of the stick hydrodynamic condition has been experimentally observed
under stress.\cite{PhysRevLett_75_2698,PhysRevLett_76_467}
Adhesion or friction occurs and the conformations of macromolecules near the interface are 
often different from the bulk. After cooling and solidifying this may influence the properties of the product
surface.\cite{ColloidPolymSci_256_1155}
Thus, it is critical to understand the rates of adsorption/desorption phenomena taking place at the interface between 
a solid and a polymeric melt.

In addition to the study of polymer melt dynamics near flat walls, there is a growing interest in the direction of 
studying nanoparticles dispersed in a polymeric matrix. This is ultimately related to the industrial importance of 
nanocomposites, in which the filler particles may have dimensions comparable to those of polymeric 
chains.\cite{AnnuRevChemBiomolEng_1_37}
Several simulation studies have appeared recently in which static and dynamic properties of model nanocomposites
were investigated.\cite{EurPolymJ_47_699, Macromolecules_46_4670, Macromolecules_47_387,Macromolecules_45_7282,
JChemPhys_140_054908, JChemPhys_140_114903}
Alongside with molecular simulations, mesoscopic models invoking the picture of ``glassy bridge'' formation around 
the dispersed nanoparticles have been developed.\cite{Macromolecules_41_8252,JPhysChemB_117_12632}
Models of this kind assume that the polymer dynamics around the dispersed particles are orders of magnitude slower
that in the melt. 
The ``percolation of free volume distribution'' model\cite{EurPhysJE_4_371} was used by Berriot and co-workers
\cite{Macromolecules_35_9756,Polymer_43_6131} for explaining the microscopic origin of the reinforcement in filled 
elastomers, as a consequence of the presence of a gradient in glass transition temperature around the fillers.
Estimates of polymer adsorption/desorption rates at the solid surface may be of help in parametrizing these models in 
a bottom-up approach.

Frantz and Granick\cite{PhysRevLett_66_899} were the first to study experimentally the adsorption-desorption kinetics 
of polymer chains from solution to a solid surface.
They monitored, by in situ Fourier-transform infrared spectroscopy, 
the surface coverage of polystyrene adsorbing to oxidized silicon from cyclohexane.
These authors have distinguished three time scales: for diffusion to and adsorption onto an initially bare surface 
(which is a rapid procedure), for ensuing surface rearrangements (slow and history dependent), and for displacement. 
The duration of the last part (i.e. the displacement of adsorbed chains along the solid surface) 
grows exponentially with the molecular weight of polymeric chains. 
Even at overall equilibrium there is a steady-state traffic of macromolecules between the adsorbed and the unadsorbed
state.\cite{JPolymSciPolymPhysEd_23_1997,Macromolecules_22_2677}

Despite the fact that plenty of simulations have been conducted on polymer/solid systems, only few of them have 
quantified adsorption/desorption kinetics.
Smith et al.\cite{Macromolecules_38_571} have carried out molecular dynamics (MD) simulations of a coarse-grained 
bead-spring model in order to examine the static and dynamic properties of polymer chains in a
melt in the presence of a solid surface. These authors have shown that the population of chains contains a subset of
weakly adsorbed chains with rapid desorption dynamics and strongly adsorbed chains with slower desorption dynamics.
The strongly adsorbed chains were held responsible for a slow-down of dynamics near the surface. The desorption time
was found to scale slightly more slowly than $N^2$ (exponent 1.5 - 2) with $N$ being the length of the chain. 
The observation holds even for chains that are in the reptation regime (where $N^3$ scaling could be expected). 
If only simple diffusion of the adsorbed beads 
were taking place, a kinetic model employed in that study would yield a desorption time scaling as $N^2$.
The MD results indicate a more complex behavior of chains that does not depend only on the number of adsorbed segments,
but probably also on the time span for which a chain has been attached to the surface (which is inversely 
related to the distance of the chain center of mass from the surface).

Yelash et al. \cite{PhysRevE_82_050801} have employed MD simulations of a chemically realistic 
model of 1,4-polybutadiene between graphite walls. They have shown that attractive confining walls introduce one more 
mechanism (in addition to the caging effect and intramolecular conformational barriers) for time scale separation of glass forming 
polymers. This mechanism is the slow desorption kinetics of monomers from the surface, leading to slow layer exchange 
dynamics on the length scale of (at least) the radius of gyration of the chains.
A layer-wise analysis of the relaxation times, as a function of the distance from the solid surface, has revealed that 
the adsorption/desorption kinetics of monomers at the wall,\cite{Polymer_51_129,Macromolecules_38_571} give rise to a 
two-step relaxation behavior, as the other two mechanisms also do. In a later study,\cite{EurophysLett_98_28006}
they have shown that a three-step (one for every timescale separation mechanism) decay can be observed in incoherent 
scattering experiments and discussed its relevance for the glass transition of confined polymers by analogy to 
surface critical phenomena.

One of the most common approaches used for the study of equilibrium properties of polymer/solid interfacial systems 
is the self-consistent field (SCF) method.\cite{Fredrickson_Book}
The essence of the SCF method\cite{JPhysCondensMatter_10_8105} is the
replacement of the ensemble of interacting polymer chains with a system of noninteracting chains subject to some
position-dependent (complex) chemical potential fields. These fields dictate the conformations of the polymer molecules, 
imposing the spatial distribution of the polymer. On the other side, the chemical potential fields depend on the 
polymer density distribution. Thus, the target of this approach is to determine the fields in such a way that they are
consistent with the spatial distribution of polymer they create.
A very important step in solving the SCF problem is the identification of stationary field configurations that
correspond to extrema of a complex effective Hamiltonian.\cite{amit1984field,Macromolecules_35_16}
For a real field theory, such a configuration can be a local minimum, maximum, or a saddle point in the field 
configuration space. However, when the field variables are extended to the complex plane, the energy surface is 
generally saddle shaped in the vicinity of an extremum, so we shall refer to stationary field configurations as 
saddle points.
It has been proved that there is an explicit connection between the saddle point chemical potential field configuration
of the SCF problem and a corresponding mean-field approximation to the free 
energy.\cite{JChemPhys_62_999,Macromolecules_14_727,amit1984field,Macromolecules_35_16}

There are two general classes of techniques that have been applied to solving SCF equations and hence to the numerical
computation of saddle points.\cite{Macromolecules_35_16}
The first are spectral methods, which attempt to represent the various spatially varying fields in a truncated 
Fourier-like basis. 
Alternatively, the nonlinear equations can be tackled in real space by suitable finite difference or 
finite element discretization of a computational domain.\cite{Macromolecules_38_7134}
Daoulas et al.\cite{Macromolecules_38_7134} have determined the equilibrium properties of a polymer melt of specific
chemical constitution (polyethylene), adsorbed on a certain solid substrate (graphite), through a continuum SCF approach.
While the monomer-monomer interactions were introduced in the simplified Helfand 
approximation,\cite{Macromolecules_8_552,JChemPhys_62_1327,Macromolecules_11_960} the interaction potential between
the polymer and the substrate was tuned in order to reproduce the total energy of adsorption and the characteristic
length scale of the density variations of the atomistically simulated system.\cite{Macromolecules_38_5780}
In contrast to previous SCF studies, for the first time, the chain connectivity was represented through the 
wormlike chain model, in addition to the simpler Gaussian model. 
Polymeric chains were found significantly ``flattened'' near the boundary, while perturbations in their
conformational properties due to the surface persisted over a characteristic length which is roughly 1.5 times the 
unperturbed mean radius of gyration. The conformations of adsorbed molecules were further characterized 
by considering the properties of tails, loops and trains, as derived by the SCF analysis of the two models (wormlike and
Gaussian) and the atomistic simulations. SCF predictions from the wormlike chain model were found closer to
the atomistic simulation results for short chain systems. The superior performance of the semiflexible chain model 
when compared to the Gaussian
one, is due to the introduction of an additional length scale, that of the persistence length. However, as the 
molecular length increases, the effect of local chain structure on global chain conformational behavior decreases,
rendering the differences between the two models insignificant.

In a recent study, Klushin et al.\cite{PhysRevE_87_022604} have investigated the effects of the range of 
adsorption potential on the equilibrium behavior of a single polymer chain end-attached to a solid surface. 
The exact analytical theory for ideal lattice chains interacting with a planar surface via a square-well potential
was presented and compared to continuum model results and to Monte Carlo (MC) simulations using the pruned-enriched 
Rosenbluth method for self-avoiding chains on a simple cubic lattice. 
We are particularly interested in the analytical approach, where the partition function of a chain with both its ends
fixed in space is obtained as the solution of the partial differential equation of diffusion of the Green's function,
an argument we will also invoke in our methodology.
In that framework, the adsorption order parameter (the bound fraction of monomers) was found to be a function of a 
single argument, the potential well depth.
Tail, loop, and train distributions at the critical point were evaluated by MC simulations and 
compared to analytical results for ideal chains and with scaling theory predictions. 
The behavior of a self-avoiding chain was found to be remarkably close to that of an ideal chain in several aspects.

The present work is aimed to address the kinetics of adsorption and desorption of an end-constrained polymer strand
belonging to a melt close to an adsorbing solid substrate.
In real-life polymer systems, end constraints may be created by crosslinks or entanglements with other chains and may 
fluctuate thermally. Such thermal fluctuations of the end positions are not considered here, in order to facilitate a 
clean mathematical formulation. The ends of the constrained chain, which in practical situations may be a subchain of 
a longer chain or of a three-dimensional network, are considered as fixed in space. Fluctuations in the chain ends can 
be introduced through a straightforward generalization of the formalism presented herein.
We employ the ``Gaussian thread model'', which considers the polymeric chains as fully flexible and infinitely 
extensible, continuous threads or ``paths'' and offers the advantage of facilitating the analytical or numerical 
evaluation of the chain conformations through the solution of the Edwards equation in the mean field 
approximation.\cite{theodorou_in_dunweg}
The fields obtained from the solution of the SCF problem are then used in order to calculate the partition function 
of end-constrained chains, as a function of the distance of their ends from the solid surface. 
This allows us to define a potential of mean force experienced by an end-constrained chain as a function of its
distance of closest approach to the solid surface. The features of the one-dimensional free energy profile are incorporated in a 
dynamically-corrected transition state theory (TST) formalism in order to estimate adsorption and desorption rate constants,
as functions of the molecular weight of the chain and the distances of its ends from the substrate.
These rates have been validated against rates obtained by hazard-plot analysis\cite{JPhysChemB_112_10619} of MD 
trajectories of the same system.\cite{atomisticMD}
The knowledge obtained from this study will be of great value for developing mesoscopic simulation approaches to treat
rate-dependent deformation and flow near solid walls or in the presence of nanoparticles. 
We expect that the characteristic time associated with exchange dynamics (adsorption/desorption) of chain strands from
the melt to the surface may play an important role in the enhancement of the mechanical properties of filled 
elastomers.

The system considered in this work is a polyethylene (PE) melt confined between two graphite phases. 
PE/graphite nanocomposites are of paramount industrial importance because they exhibit electrical and thermal 
conductivity\cite{SyntMet_145_245}, holding a great promise as antistatic 
materials, capable of dissipating static charges safely from component surfaces.\cite{JVinylAdditTechnol_19_258}
The thermal conductivity of PE/graphite composites can be as high as $11.28 \;{\rm W \:m}^{-1}{\rm K}^{-1}$.
\cite{MaterDes_52_621} 
These composites exhibit a temperature-dependent electrical conductivity due to thermal expansion-dependent 
percolation of the carbon particles.\cite{ElectronMaterLett_7_249,JMaterSci_32_401,JMaterSci_32_1711}
The electrical resistivity of carbon black filled with high density polyethylene increases significantly when the 
composite is heated to the melting temperature of the matrix.\cite{Carbon_39_375}
However, void formation was observed to occur during polyethylene crystallization as a result of filler 
particle/polymer matrix interactions, by small angle neutron scattering.\cite{Macromolecules_32_5399}
Moreover, PE is not too different from cis-1,4 polyisoprene and the surface of carbon black could be approximated as 
graphite, so our system is close, at a molecular level, to carbon black-reinforced rubbers.

\section{SCF Formulation for a Homopolymer Melt Next to an Adsorbing Solid Surface}
\subsection{Field Theoretic Formulation of the Grand Partition Function}
The continuum SCF approach is based on a path-integral representation of the partition function. In the past, this 
approach has been extensively employed in numerous 
works.\cite{JPhysCondensMatter_10_8105,Macromolecules_35_16,AdvPolymSci_185_1}
Here, an effort will be made to present the theory as briefly as possible, emphasizing only essential points
specific to the present work. We consider a polymer melt in the presence of a solid phase, which exerts a 
position-dependent field $U_{\rm s}\left(\mathbf{r}\right)$ on the polymer segments. Each chain is envisioned as
a sequence of $N$ statistical segments obeying Gaussian statistics, and $U_{\rm s}\left(\mathbf{r}\right)$ is an 
energy per statistical segment located at $\mathbf{r}$. We focus on a region of total volume $V$ occupied by polymer
at temperature $T$. The boundaries of that region are partly defined by the surface(s) of the solid. The polymer in the
considered control volume is at equilibrium with a bulk polymer phase, with which it can freely exchange chains. The 
chemical potential of a chain in that phase is $\mu N$ (where by definition, $\mu$ is the chemical potential per 
segment). The polymer in the interfacial region conforms to the probability distribution of the grand canonical ensemble.
Following the statistical mechanical development of Daoulas et al.,\cite{Macromolecules_38_7134} based on a simplified,
Helfand-type effective Hamiltonian for polymer-polymer interactions, the grand canonical partition function can be 
written as in Eq.(2) of that paper:
\begin{align}
\Xi = & \mathlarger{\sum}_{n=0}^\infty \frac{1}{n!} \exp{\left(\frac{\mu N n}{k_{\rm B}T} \right)} \tilde{N}^n
\mathlarger{\int} \prod_{\alpha = 1}^{n} \mathcal{D} \mathbf{r}_{\alpha}\left( \bullet \right) 
\mathcal{P} \left[\mathbf{r}_{\alpha}\left( \bullet \right) \right] \nonumber \\
& \exp\left( -\frac{1}{2 \kappa_{\rm T} k_{\rm B} T} \mathlarger{\int} \left[\hat{\phi}\left(\mathbf{r} \right) - 1 \right]^2
\D^3 r \right . \nonumber \\
& \;\;\;\;\;\;\;\;\; \left . - \frac{\rho_0}{k_{\rm B}T} \mathlarger{\int}U_{\rm s} \left(\mathbf{r}\right) 
\hat{\phi} \left(\mathbf{r}\right)
\D^3 r \right) 
\label{eq_grand_partition_function}
\end{align}
where $k_{\rm B}$ is the Boltzmann constant,
$\rho_0$ and $\kappa_{\rm T}$ are the mean segment density and isothermal compressibility in the bulk melt,
while $\mu$, $N$, and $n$ are the segment chemical potential, the chain 
length of the coarse-grained Gaussian thread, and the number of chains, respectively. $\tilde{N}$ is a normalizing 
prefactor, and $\mathcal{D} \mathbf{r}_{\alpha} \left(\bullet \right)$ denotes functional integration over all possible 
conformations (``paths'') of chain $\alpha$, while $\hat{\phi}\left(\mathbf{r} \right)$ is the volume fraction 
operator, given by
\begin{equation}
\hat{\phi}\left(\mathbf{r}\right) = \frac{N \mathlarger{\sum}_{\alpha =1}^n 
\mathlarger{\int}_0^1 \delta\left(\mathbf{r} - \mathbf{r}_\alpha (s) \right) \D s}{\rho_0}
\label{eq_phihat_definition}
\end{equation}
where $s$ is a scaled variable measuring how far along the contour length of the chain a considered segment lies; 
it ranges from $0$ (chain start) to $1$ (chain end). 
The functional $\mathcal{P} \left[\mathbf{r}_{\alpha}\left( \bullet \right) \right]$ accounts for the chain 
connectivity, and in the case of the Gaussian model it takes the form
\begin{equation}
\mathcal{P} \left[\mathbf{r}_{\alpha}\left( \bullet \right) \right]  = \exp{\left[ -\frac{1}{4 R_{\rm g}^{2}} 
\mathlarger \int_{0}^{1} \left(\frac{\D \mathbf{r}}{\D s} \right)^{2} \D s \right]}	
\end{equation}
where $R_{\rm g}^{2}$ is the mean-squared radius of gyration of the polymer chain in the bulk. It can be seen that,
in the case of the Gaussian model, only one characteristic conformational parameter of the atomistic polymer chain
enters the coarse grained model: $R_{\rm g}^{2}$. 

Following a standard field theoretical approach, it is possible to replace the system of interacting chains subject to
the field of the solid with a system of noninteracting chains subject to the field of the solid and an additional 
fluctuating field representing the remaining chains. 
This procedure is described in detail in Appendix A. We introduce the notation $Q\left[i w + U_{\rm s}\right]$ to 
indicate the partition function of a single chain subject to the field $iw + U_{\rm s}$ acting on its segments, 
relative to the partition function of a field-free chain. 
The grand partition function can be written as:
\begin{equation}
\Xi = C \mathlarger \int \mathcal{D} \left[ \beta w \right] \exp{\left\{ - \beta H\left[w\right] \right\}}
\label{eq_grand_partition_function_effective_hamiltonian}
\end{equation}
with
\begin{align}
   H\left[ w \right]  = & \mathlarger \int \D^{3}r \left[-i \rho_{0} w\left(\mathbf{r}\right) 
   + \frac{\kappa_{\rm T}}{2} \left(\rho_{0} w\left(\mathbf{r}\right) \right)^{2} \right]
   \nonumber \\
   - & \frac{1}{\beta} \exp{\left(\frac{\mu N}{k_{\rm B}T} \right)} \tilde{N} Z_{\rm free} Q\left[iw + U_{\rm s}\right]
   \label{eq_eff_hamiltonian}
\end{align}
It is remarkable that the the effective ``Hamiltonian'' $H\left[w\right]$ incorporates a term proportional to 
$\exp{\left(\frac{\mu N}{k_{\rm B}T} \right)}$ and to $Q$, rather than to $\mu N$ and to $\ln{Q}$, because of the grand
canonical formulation adopted.
$Z_{\rm free}=\int \mathcal{D}\mathbf{r}_\alpha \left(\bullet\right) 
\mathcal{P}\left[\mathbf{r}_\alpha\left(\bullet\right)\right]$ is the partition function of a free chain.
By invoking a saddle point approximation (which is described in detail in Appendix B), the
effective Hamiltonian,  \ref{eq_eff_hamiltonian}, at the optimum can be written as:
\begin{align}
   \bar{H} =  - & \rho_0 \mathlarger \int \mathrm{d}^3 r w^\prime \left(\mathbf r \right) 
                 \phi \left(\mathbf r \right) 
           +  \frac{1}{2\kappa_{\rm T}} \mathlarger \int \mathrm{d}^3 r \left[1 - \phi \left(\mathbf r \right) \right]^2
                  \nonumber \\
           + & \rho_0 \mathlarger \int \mathrm{d}^3 r U_{\rm s} \left(\mathbf r \right) 
                  \phi \left(\mathbf r \right) \nonumber \\
           - & \frac{1}{\beta} \exp{\left(\frac{\mu N}{k_{\rm B}T} \right)} \tilde{N} Z_{\rm free} Q
           \left[\imun w + U_{\rm s} \right] \label{eq_eff_Hamiltonian_optimum}
\end{align}
where we have set 
$w^\prime \left(\mathbf r \right) = \imun w \left(\mathbf r \right) + U_{\rm s} \left( \mathbf r \right)$,
a real field (see Appendix A). 
The effective field $w^\prime \left(\mathbf{r} \right)$ and the volume fraction, 
$\phi \left( \mathbf r \right)$, are coupled to each other through a set of equations depending on the model adopted for the
chain connectivity.\cite{Macromolecules_38_7134}
For the case of the Gaussian threads considered in this work, assuming that distances between solid surfaces present
are large in comparison to $R_{\rm g}$, such that bulk conditions prevail far from the surfaces in the interfacial 
polymer, these equations are formulated as:
\begin{equation}
   w^\prime \left(\mathbf{r} \right) = \frac{1}{\kappa_{\rm T} \rho_0} \left[\phi\left(\mathbf r \right) - 1 \right]
   + U_{\rm s} \left(\mathbf r \right)
   \label{eq_sqf_real_field_definition}
\end{equation}
which is the self-consistent field expression for the real field and 
\begin{equation}
   \phi\left(\mathbf r \right) = \int_0^1 \mathrm{d}s\; q \left({\mathbf r},s\right) q \left({\mathbf r}, 1-s \right)
   \label{eq_segment_balance_condition}
\end{equation}
which stands for the segment balance condition. 
In the formulation adopted here, $\phi \left({\mathbf r}\right)$ is a segment density reduced by its value $\rho_0$ in 
the bulk melt at the same polymer chemical potential. 
The quantity $q \left({\mathbf r},s\right)$ is the restricted partition function, proportional to the probability 
density that the segment at fractional contour length $s$ of a chain which may have started anywhere in the system 
lies at position $\mathbf{r}$. It is normalized by the corresponding probability density in the bulk melt and obeys 
the initial condition $q\left({\mathbf r}, 0 \right) = 1$. The restricted partition function 
$q\left({\mathbf r}, s \right)$ can be calculated through the Edwards diffusion equation:
\begin{equation}
   \frac{\partial q\left(\mathbf r ,s \right)}{\partial s} = R_{\rm g}^2 \nabla_{\mathbf r}^2 q\left(\mathbf r ,s \right)
   - \beta N w^\prime \left(\mathbf r \right) q\left(\mathbf r ,s \right)
   \label{eq_edwards_diffusion_equation}
\end{equation}
A relation exists between the restricted partition function $q\left(\mathbf r ,s \right)$ and the single chain 
partition function in the presence of the field, $Q\left[\imun w + U_{\rm s} \right] \equiv Q\left[w^\prime \right]$:
\begin{equation}
   Q \left[w^\prime \right] = \frac{1}{V} \mathlarger \int {\rm d}^3 r \: q\left(\mathbf r, 1\right)
\end{equation}

The SCF problem to be solved in order to capture structure and thermodynamics in the interfacial system consists of the
partial differential ``diffusion'' equation,  \ref{eq_edwards_diffusion_equation}, the definition of the the self-consistent
field,  \ref{eq_sqf_real_field_definition}, and the segment balance  \ref{eq_segment_balance_condition}.
These must be solved numerically in the unknown restricted partition function $q\left({\mathbf r},s\right)$, the 
volume fraction profile $\phi \left({\mathbf r}\right)$, and the self-consistent field $w^\prime \left( {\mathbf r}\right)$.
As already pointed out, the ``initial'' condition $q\left({\mathbf r},0\right) = 1$ 
applies.\cite{PhysRevLett_73_3235,JChemPhys_104_7758,Macromolecules_35_16}
Additionally, boundary conditions must be used that reflect the geometry of the problem.

\subsection{Numerical Solution of the SCF Problem}

We specialize to the one-dimensional problem, where the polymeric melt is confined between two identical planar surfaces, 
within a gap of width $2 L_z$. After recalling the translational invariance of the system along the $xy$ plane,
we can write 
$q\left({\mathbf r},s\right) = q\left(z,s\right)$ and $\phi\left( {\mathbf r}\right) = \phi\left(z \right)$. 
We have to solve the following integrodifferential system
in the unknown functions $q\left(z,s\right)$, $\phi\left(z\right)$ and $W\left(z\right)$ in the domain 
$0 \le z \le L_z$ and $0 \le s \le 1$:
\begin{align}
   \frac{\partial q \left(z,s\right)}{\partial s} & = R_{\rm g}^2 \frac{\partial^2 q\left(z,s\right)}{\partial z^2}
   - W\left(z\right)q\left(z,s\right) \label{eq_1d_Edwards_diffusion_equation} \\
   \phi\left(z\right) & = \int_0^1 \D s \: q\left(z,s\right) q\left(z,1-s\right) 
   \label{eq_1d_volume_fraction}\\
   W\left(z\right) & = \frac{\beta N}{\kappa_{\rm T}\rho_0} \left[\phi\left(z\right) -1\right] 
   + \beta N U_{\rm s} \left(z\right) \label{eq_1d_real_field}
\end{align}
under the boundary conditions:
\begin{align}
   q (0,s) & = 0 \label{eq_wall_absorbing_boundary_condition}\\
   \left . \frac{\partial q \left(z,s\right)}{\partial z} \right|_{z = L_z} & = 0  
   \label{eq_plane_of_symmetry_boundary_condition}
\end{align}
and the initial condition:
\begin{equation}
   q (z,0) = 1,\;\; \forall z > 0
   \label{eq_1d_initial_q_condition}
\end{equation}
The plane of symmetry at $z = L_z$,  \ref{eq_plane_of_symmetry_boundary_condition}, implies that the polymer is 
sandwiched between two identical surfaces.
The problem is solved within half of the gap only. We are here interested in cases where the thickness $L_z$ of
the interfacial domain, in which the problem is solved, is large enough, in comparison to the range of the 
segment-surface interactions and in comparison to the chain dimensions $R_{\rm g}$, for bulk conditions to prevail
at $z = L_z$. We expect that $U_{\rm s} \left(L_z \right) =0$ and $\phi \left(L_z \right) = 1$, hence 
$W \left( L_z \right) = 0$ as well. However, it is certainly possible to address thin films of polymer, such that
bulk conditions are not realized anywhere in the film.

Commonly, \cite{Macromolecules_29_2179,JChemPhys_114_5366,Macromolecules_37_3026} when considering questions related
to polymer adsorption, the Edwards diffusion equation is supplied with an effective boundary condition. 
This practice originates from the work of de Gennes\cite{RepProgPhys_32_187} and amounts to splitting the total 
effective field felt by a polymer segment into a part originating from its interaction with the substrate and a
mean-field part accounting for the interactions with the rest of the polymer segments. In our case, however, following
Daoulas et al.,\cite{Macromolecules_38_7134} we have to preserve the surface component of the effective field, 
$W\left(z\right)$. To this end, the propagator equation,  \ref{eq_1d_Edwards_diffusion_equation}, has to be supplied
with an absorbing boundary condition,  \ref{eq_wall_absorbing_boundary_condition}, 
requiring $q \left(z = 0, s\right) = 0$.
Wu et al.\cite{JPolymSciPartBPolymPhys_33_2373} gave a detailed reasoning of how, beyond some distance from an 
enthalpically neutral substrate, the microscopic absorbing boundary condition effectively reduces to the well-known 
for this case reflecting boundary condition.

The SCF equations formulated in the current work (i.e. eqs \ref{eq_1d_Edwards_diffusion_equation}, 
\ref{eq_1d_volume_fraction} and \ref{eq_1d_real_field}) can be solved following a Newton-Raphson method, which is 
expected to converge fast, albeit, requiring more memory and CPU time per iteration than the more commonly used 
successive substitution, due to the need to compute and store the Jacobian. 
The details of the discretization and numerical solution of the system of eqs \ref{eq_1d_Edwards_diffusion_equation} 
-\ref{eq_1d_initial_q_condition} can be found in the Supporting Information to the present paper.

\section{SCF-Based Formulation for Estimating the Rates of Adsorption and Desorption of Entangled or Crosslinked
Strands from a Melt or Rubbery Polymer}

\subsection{Formulation of Adsorption and Desorption as Infrequent Event Processes}
We consider a polymer above its glass transition temperature, consisting of randomly coiled strands whose ends are more
or less constrained by entanglements or crosslinks, next to a flat solid surface exerting a potential energy field
$U_{\rm s}\left(\mathbf{r}\right)$ on polymer segments. Our objective is to quantify the rates
of adsorption and desorption of a strand in the sea of other strands, as functions of the 
positions of the two ends ${\mathbf r}\left(s = 0\right)$ and ${\mathbf r}\left(s = 1\right)$, with $s$ being the 
reduced contour length along the strand.  
In the following we will use the terms ``strand'' and ``chain'' interchangeably, even though the considered strand 
may in fact be a subchain of a longer chain.
Without loss of generality, we consider within a linear monodisperse melt a chain of length $N$ statistical segments
whose ends are constrained at or near ${\mathbf r}(0)$ and ${\mathbf r}(1)$. 
The conformations of such a chain in the interfacial region can be analyzed via a SCF model of the type we discussed
above. This model provides probability distributions for the ends ($s=0$ and $s=1$) to find themselves in specific
positions in space.

\begin{figure}
   \centering
   \includegraphics[width=0.45\textwidth]{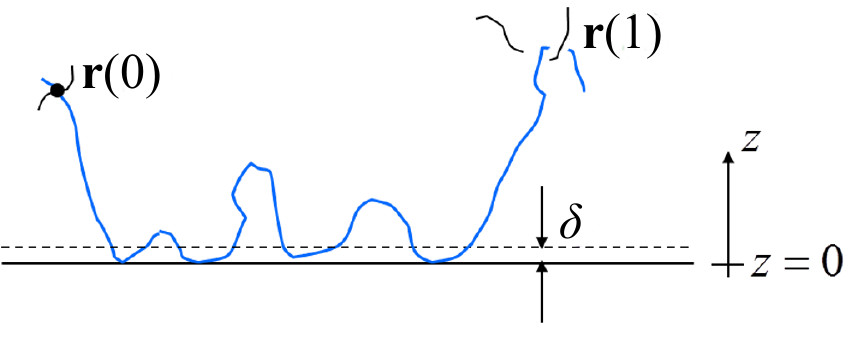}
   \caption{Schematic representation of an adsorbed chain. In this example, one of the chain ends, ${\mathbf r}(0)$, 
   is a crosslink, while the other one, ${\mathbf r}(1)$, is an entanglement point with another chain 
   (marked in black). The height of the layer, in which adsorption is considered to take place, is $\delta$.}
   \label{fig:tst_schematic}
\end{figure}

Based on the form of the segment-surface interactions, $U_{\rm s}(z)$, one can define a thickness $\delta$ next
to the surface such that, if a segment lies at $ 0 < z <\delta$ with $z = 0$ indicating the plane of the surface,
it is considered as adsorbed, while if it lies at $z \ge \delta$, it is considered as free. A chain will be 
considered as adsorbed (subscript ``a'') if it contains at least one adsorbed segment. Otherwise it will be considered
as free (subscript ``f''). An adsorbed chain, lying between a crosslink and an entanglement, is depicted in 
Figure \ref{fig:tst_schematic}, along with the basic geometric quantities used in the formulation.
The chain may execute a very complex motion near the surface, involving fluctuations in the number and length of
trains, loops and tails.\cite{Macromolecules_38_571} In this work, we intend to derive overall rate constants for 
the rates of adsorption and desorption of the end-constrained chain on the surface, which are applicable to the long-time
behavior of the chain and are integrated over all these fluctuations. With the chain ends fixed at ${\mathbf r}(0)$ and 
${\mathbf r}(1)$ in the $z > \delta$ domain, we think of adsorption as a transition between a ``free'' state 
($z(s) \ge \delta$) and an ``adsorbed'' state ($z(s) \le \delta$ for some $s$ in $0 \le s \le 1$).

\subsection{Configurational Integral of End-constrained Chains}
In the context of the continuous SCF model of a melt next to an adsorbing solid surface, we define 
$\mathcal{Z}(z)$ as a configurational integral over all chains, with their ends fixed at ${\mathbf r}(0)$ and 
${\mathbf r}(1)$, which do not have any segments in the region $0 < z^\prime < z$, with $z^\prime = 0$ denoting 
the substrate surface and $z^\prime = \delta$ denoting the dividing surface for adsorption. $\mathcal{Z}(z)$ is 
expected to be a decreasing function of $z$.

For the calculation of $\mathcal{Z}(z)$ we define a propagator 
$G_{>z}\left({\mathbf r}_i, {\mathbf r}^{\prime}, s \right)$ for a 
chain that has started at position ${\mathbf r}_i$ to be at position ${\mathbf r}^{\prime}$ at contour length $s$, 
\emph{the whole contour of the chain from 0 to $s$ lying entirely in the region of space $z^{\prime} > z$}. 
This is equivalent to posing an imaginary neutral solid wall at $z$, while solving the diffusion
differential equation,  \ref{eq_propagator_diffusion}.
We expect $G_{>z}\left({\mathbf r}_i, {\mathbf r}^{\prime}, s \right)$ to be factorizable,
due to the infinite extensibility of the Gaussian chain model. Presumably, for finite-contour length 
models, like the freely-jointed or the worm-like chain models, such a factorization is not possible.
Thus, the propagator $G_{>z}\left({\mathbf r}_i, {\mathbf r}^{\prime}, s \right)$ becomes:
\begin{equation}
   G_{>z}\left({\mathbf r}_i, {\mathbf r}^{\prime}, s \right) = 
   G_{x}\left(x_i, x^{\prime}, s \right)
   G_{y}\left(y_i, y^{\prime}, s \right) 
   G_{>z} \left(z_i, z^{\prime}, s\right)
\end{equation}
with the factors $G_{x}\left(x_i, x, s\right)$, $G_{y}\left(y_i, y, s \right)$ being unconstrained 
one-dimensional Gaussian functions emanating from $x_i$, $y_i$, respectively, and therefore independent 
of $z$. By definition, 
\begin{equation}
   G_{>z} \left(z_i, z^{\prime}, s \right) = 0 \;\; \forall z^{\prime} \;\text{ for } z>z_{i}
\end{equation}
so $G_{>z}\left(z_i,z^{\prime},s \right)$ should be computed for $0 < z < z_{i}$ and $z^{\prime} > z$.

The configurational integral $Z(z)$ is related to $G_{>z}\left(z_i, z^{\prime},s\right)$ as follows:
\begin{align}
   \frac{\mathcal{Z}(z)}{\mathcal{Z}_{\rm bulk}} = &
   \int_{z^{\prime}>z} \D z^{\prime} \int_{-\infty}^{+\infty} \D x \int_{-\infty}^{+\infty} \D y  
   \int_{0}^{1} \D s \nonumber \\
   & \left [ G_{>z}\left(z(0), z^{\prime},s\right) 
         G_{>z}\left(z(1), z^{\prime},1-s\right) \right . \nonumber \\ 
   & \times G_{x}\left(x(0), x,s \right) G_{x}\left(x(1), x, 1-s\right)
   \nonumber \\
   & \left . \times G_{y}\left(y(0), y,s \right) 
   G_{y}\left(y(1), y, 1-s\right)\right ]
\label{eq_z_from_g_integrals}
\end{align}
where $\mathcal{Z}_{\rm bulk}$ is the configurational partition function of a free chain of length $N$ starting at 
a given point in the melt.
It should be noted that $\mathcal{Z}_{\rm bulk}$ is independent of the exact starting point of the free chain.
Note also that the ratio $\mathcal{Z}(z)/\mathcal{Z}_{\rm bulk}$ has dimensions of inverse volume, 
since $\mathcal{Z}_{\rm bulk}$ in the denominator has only one end constrained, while $\mathcal{Z}(z)$ has two 
ends constrained. 
The right-hand side of  \ref{eq_z_from_g_integrals} also has dimensions of inverse volume, since the propagators 
along all $x$, $y$ and $z$ have dimensions of inverse length.
The propagator $G_{>z}\left(z_i, z^{\prime},s\right)$ needed within  \ref{eq_z_from_g_integrals} for
$z_i = z(0)$ and $z_i = z(1)$ satisfies the following differential equation:
\begin{align}
   \frac{\partial G_{>z}\left(z_i, z^{\prime},s\right)}{\partial s} & = R_{\rm g}^2 
   \frac{\partial^2 G_{>z}\left(z_i, z^{\prime},s\right)}{\partial z^{\prime 2}} \nonumber \\
   & - W(z^\prime) G_{>z}\left(z_i, z^{\prime},s\right), \; s > 0  \nonumber \\
   G_{>z}\left(z_i, z^{\prime}, 0\right) & = \delta \left(z^\prime - z_i\right) 
   \text{ (initial condition) } \nonumber \\
   G_{>z}\left(z_i, z, 0\right) & = 0 
   \text{ (boundary condition at } z^\prime = z \text{ ) } \nonumber \\
   \lim_{z^\prime \to \infty} G_{>z}\left(z_i, z^\prime, s\right)  &= 0 
   \text{ (boundary condition at } z^\prime \to \infty { )}
   \label{eq_propagator_diffusion}
\end{align}
The absorbing boundary condition at $z^\prime = z$ stems from the requirement that the entire contour of the considered
chain must lie at $z^\prime > z$. The self-consistent field $W\left(z^\prime\right)$ is already known from our 
solution to the SCF problem. Eq \ref{eq_propagator_diffusion} must be solved for given $z_i$, 
$z$ with $0<z<z_i$ to determine $G_{>z}\left(z_i, z^{\prime},s\right)$ for all $z^\prime > z$.

For our one-dimensional problem, the integrations over $x$ and $y$ appearing in  \ref{eq_z_from_g_integrals} can be
carried out analytically, yielding:
\begin{align}
   & \int_{-\infty}^{+\infty} \D x^\prime \: G_{x}\left(x(0), x^{\prime},s \right) G_{x}\left(x(1), x^{\prime}, 1-s\right)
   \nonumber \\
   = &  \frac{1}{2 \sqrt{\pi} R_{\rm g}} \exp{\left[ -\frac{\left[x(1) - x(0) \right]^2}{4 R_{\rm g}^2} \right]}
   \label{eq_propagator_along_x}   
\end{align}
independent of $s$, as one would expect, since we have an unconstrained Gaussian chain in the $x$-direction.
Eq \ref{eq_propagator_along_x} is readily applicable in the $y$-direction, too. 
From eqs \ref{eq_z_from_g_integrals} and \ref{eq_propagator_along_x}:
\begin{align}
   \frac{\mathcal{Z}(z)}{\mathcal{Z}_{\rm bulk}} = & \frac{1}{4 \pi R_{\rm g}^2} \exp{\left[
   -\frac{\left[x(1) - x(0)\right]^2 + \left[y(1) - y(0) \right]^2}{4 R_{\rm g}^2} \right]} \times \nonumber \\
   & \int_{z^{\prime}>z} \D z^{\prime} \int_{0}^{1} \D s \;
   G_{>z}\left(z(0), z^{\prime},s\right)
   G_{>z}\left(z(1), z^{\prime},1-s\right) 
   \label{eq_z_final_from_g_integrals}
\end{align}

Let $\mathcal{Z}_{\rm a}$ and $\mathcal{Z}_{\rm f}$ be the configurational integrals of the chain in the adsorbed and
free states, for given ${\mathbf r}(0)$ and ${\mathbf r}(1)$. The former one can be cast as the difference between 
the configurational integral at $z=0$ (all chains having their segments closest to the surface above the $z=0$ level) 
and the configurational integral at $z=\delta$ (all chains having their segments closest to the surface above the
$z = \delta$ level). With these definitions: 
\begin{equation}
   \mathcal{Z}_{\rm f} = \mathcal{Z} \left( \delta \right) ,\;
   \mathcal{Z}_{\rm a} = \mathcal{Z} \left( 0 \right) - \mathcal{Z}\left( \delta \right) 
\end{equation}
and the equilibrium probabilities $P_{\rm f}$ and $P_{\rm a}$ of the chain being free and adsorbed, respectively, are:
\begin{equation}
   P_{\rm f} = \frac{\mathcal{Z}_{\rm f}}{\mathcal{Z}_{\rm f} + \mathcal{Z}{\rm a}},\; 
   P_{\rm a} = \frac{\mathcal{Z}_{\rm a}}{\mathcal{Z}_{\rm f} + \mathcal{Z}{\rm a}} 
   \label{eq_equilibrium_probabilities}
\end{equation}

\subsection{Connection of Infrequent Event Formulation with SCF Model}

Clearly, associated with the configurational integral $\mathcal{Z} (z)$ is a potential of mean force 
(configurational free energy) for a chain to have its segment(s) lying closest to the surface at $z$.
We envision that, in order to desorb from the adsorbed state or adsorb from the free state, the chain must pass
through a ``transition state'' configuration that has at least one segment in the plane $z = \delta$ but all other
segments at $z > \delta$. In other words, we consider as ``reaction coordinate'' the lowest among all $z$-coordinates 
of the segments of the considered chain and a ``dividing surface'' for this reaction coordinate at $z = \delta$.
By definition, then, $- \D \mathcal{Z}(z)/\D z$ is a configurational integral per unit length along the 
$z$-direction for a chain, with its ends fixed at ${\mathbf r}(0)$ and ${\mathbf r}(1)$, to have its segment(s) 
closest to the surface between $z$ and $z+\D z$.
Thus, the \emph{configurational integral per unit length} at the dividing surface is:
\begin{equation}
   \mathcal{Z}^\dagger = - \left. \frac{\D \mathcal{Z}}{\D z}\right|_{z = \delta}
\end{equation}
while the potential of mean force can be obtained as:
\begin{equation}
   A \left(z\right) = - k_{\rm B} T \ln{\left[- \frac{\D \mathcal{Z}(z)}{\D z} C_4 \right]}
   \label{eq_potential_of_mean_force}
\end{equation}
where a multiplicative constant with the units of length to the fourth power, $C_4$, is introduced in the 
potential of mean force, in order to make 
the argument of the logarithm dimensionless. The introduction of the constant shifts the reference level for the 
estimation of free energies without affecting their physical significance. We expect the potential of mean force,
$A(z)$, to exhibit a local maximum with respect to $z$, which will be used as an estimate of the position of
the dividing surface.

The next step in calculating the rates of adsorption and desorption would now be to apply dynamically corrected 
transition state theory, based on the theory of Bennet\cite{Bennet_1975} and Chandler.\cite{JChemPhys_68_2959}
It would be reasonable to assume that the chain is well described by Kramers\cite{Physica_7_284} theory in the high 
friction limit. If we consider a Rouse picture for the chain in the melt, and let $\zeta_{\rm strand}$ denote a 
friction coefficient between a strand and the rest of the melt (proportionality constant between frictional force 
exerted by the melt and velocity of the strand relative to the melt, 
measured in units of mass per time), the dynamical correction factor associated with recrossing the dividing surface 
will be:
\begin{equation}
   f = \frac{ \sqrt{\kappa_{\rm b} m_{\rm strand} }}{\zeta_{\rm strand}}
   \label{eq_correction_factor}
\end{equation}
where $\kappa_{\rm b}$ is related to the curvature of the free energy profile at the top of the barrier:
\begin{equation}
  \kappa_{\rm b} = - \left . \frac{\D^2A(z)}{\D z^2}\right|_{z = \delta}
  \label{eq_fenergy_curvature}
\end{equation}
By assuming an effective mass $m_{\rm strand}$ for a polymer strand, 
at temperature $T$ the strand will have a mean thermal speed along $z$ (mean of the magnitude of the velocity in 
one dimension, assuming a Boltzmann distribution):
\begin{equation}
\left\langle \left | v_{z}\right | \right \rangle = \left( \frac{2 k_{\rm B} T}{\pi m_{\rm strand}}\right)^{1/2}
\end{equation}
Thus, the adsorption rate constant for the end-constrained chain emerges as:
\begin{align}
   k_{{\rm f}\to{\rm a}} = & \frac{1}{2} f\left\langle \left| v_{z}\right | \right\rangle 
   \frac{\mathcal{Z}^\dagger}{\mathcal{Z}_{\rm f}} \nonumber \\
   = & \frac{1}{\zeta_{\rm strand}}
   \left(\frac{\kappa_{\rm b} k_{\rm B} T}{2 \pi} \right)^{1/2} \frac{1}{\mathcal{Z}\left(\delta\right)}
   \left( - \left . \frac{\D \mathcal{Z}}{\D z}\right|_{z = \delta} \right)
   \label{eq_adsorption_rate_constant}
\end{align}
with $\kappa_{\rm b}$ being given in terms of $\mathcal{Z}\left(z\right)$ via eqs \ref{eq_fenergy_curvature} and
\ref{eq_potential_of_mean_force}.
Similarly, the desorption rate constant for the end-constrained chain emerges as:
\begin{align}
   k_{{\rm a }\to {\rm f}} = & \frac{1}{2} f \left\langle \left| v_{z}\right | \right\rangle
   \frac{\mathcal{Z}^\dagger}{\mathcal{Z}_{\rm a}} \nonumber \\
   = & \frac{1}{\zeta_{\rm strand}}
   \left(\frac{\kappa_{\rm b} k_{\rm B} T}{2 \pi} \right)^{1/2} 
   \frac{1}{\mathcal{Z}\left(0 \right) - \mathcal{Z}\left(\delta\right)}
   \left( - \left . \frac{\D \mathcal{Z}}{\D z}\right|_{z = \delta} \right)
   \label{eq_desorption_rate_constant}
\end{align}

\section{Connection to Atomistic Simulations}

\subsection{Representation of the PE Melt/Graphite Interfacial System in the SCF/TST framework}
After the set of SCF and TST equations has been formulated, there are several parameters to be determined, so that 
a correspondence between the field-theoretic and the atomistic PE/graphite systems can be established.
Since all atomistic simulations were performed at $T = 450 \;{\rm K}$, the same temperature will be assumed during the
SCF/TST calculations of the present work. The bulk density of the polymer is $\rho_{0} = 0.766 \;{\rm g/cm}^3$.
The compressibility appearing in the SCF equations, $\kappa_{\rm T}$, will equal the isothermal compressibility of the
atomistically studied bulk polymer.\cite{atomisticMD} Taking into account our atomistic simulations and experimental
data,\cite{Polymer_33_3462} we employ $\kappa_{\rm T} = 1.43\times 10^{-9}\;{\rm Pa}^{-1}$.
The length scales of the field-theoretic and the atomistic molecular models are related through two invariant 
parameters: the contour length of the chain, $L$, and the mean squared radius of gyration of the chain, $R_{\rm g}^2$.
The latter can be calculated as $R_{\rm g}^2 = (C/6) N_{\rm b} l_{\rm b}^2$ where $C$ is the polyethylene characteristic ratio,
$N_{\rm b}$ the number of chemical bonds along the chain and $l_{\rm b}$ their bond length. The bond length used in the
atomistic simulations is $l_{\rm b} = 1.54 \;\text{\AA}$ and we set $C$  for the molecular weight of interest, 
following the work of Karayiannis et al.\cite{JChemPhys_117_5465}
A crucial step in implementing the field-theoretic representation for the PE/graphite system is the definition
of the polymer/substrate interaction potential, $U_{\rm s}(z)$. In this scope, we follow the work of Daoulas et
al.\cite{Macromolecules_38_7134} where a square-well potential was used. The depth of the square well potential 
was determined by requiring that the field-theoretic model have the same adsorption energy per unit surface as the
atomistic one. 

The Rouse model\cite{JChemPhys_21_1272} considers a polymer chain as a sequence of Brownian
particles connected by harmonic springs and moving in a viscous medium representing the background environment 
formed by all other chains. An important parameter in the Rouse formulation is the friction factor $\zeta_0$ (measured
in units of mass per time), characterizing the resistance encountered by a monomer unit moving through its 
surroundings. It can be thought of as the proportionality constant between the velocity of the 
Brownian particle and 
the frictional force exerted on the particle as it moves through the ``sea'' formed by all other particles in the system. 
Harmandaris et al.\cite{Macromolecules_31_7934} have estimated the segmental friction coefficient of polyethylene 
carbon atoms by mapping atomistically detailed MD simulations of linear PE melts onto the Rouse model. 
Their simulations have shown that there is a minimum chain length (around \ce{C60}), above which $\zeta_0$ can be 
considered as constant, chain length independent parameter of the system. The asymptotic value is 
$\zeta_0 \simeq 4.5 \times 10^{-13} \;{\rm kg/s}$.
If we think of the friction coefficient as being proportional to the mass of the entity it refers to, we can estimate the
maximum possible friction coefficient for the diffusion of the strands treated in the TST framework as: 
$\zeta_{\rm strand} = \left(m_{\rm strand} / m_{{\rm CH}_2} \right) \zeta_0$. 
The real friction coefficient of the strand will lie between the extreme values: 
$\zeta_0 < \zeta_{\rm strand} < \zeta_{0}\frac{m_{\rm strand}}{m_{{\rm CH}_2}}$.
We do not take into account the possible dependence of the friction coefficient $\zeta_0$ on the distance of the 
strand from the graphite surface.\cite{JChemPhys_140_054908}

\subsection{Atomistic Simulations}
The atomistic simulations were conducted by describing the PE chains with a fully flexible united atom model, which
considers each methylene (\ce{CH2}) and methyl (\ce{CH3}) group along the chain backbone as one interaction 
site.\cite{Macromolecules_38_5780,Macromolecules_38_5796}
Polymer intramolecular interactions (bond stretching, bond angle bending, and torsional potential) are implemented 
following Nath et al.\cite{JChemPhys_108_9905}
Non-bonded interactions between pairs of \ce{CH2} or \ce{CH3} are described by a pairwise-additive Lennard-Jones 
12-6 potential with input parameters provided by the TraPPE forcefield of Martin and Siepmann.\cite{JPhysChemB_102_2569}
Ewald summation has been applied to the attractive part of all polymer-polymer nonbonded 
interactions,\cite{JPhysChem_93_7320,JChemPhys_127_144711}
in order to avoid distortions of the density profile and associated stress profile.
To calculate the potential energy of interaction between a polymer atom and a semiinfinite graphite phase, we implement
an efficient and accurate Fourier summation method designed by Steele,\cite{SurfSci_36_317} which takes advantage of 
the symmetry in the crystalline substrate. 
For more details and a thorough study by using atomistic simulations the interested reader is referred to our 
accompanying work on the same system.\cite{atomisticMD}
In the following paragraphs we will summarize the vital pieces of information relative to the present work.

Initial configurations of polyethylene confined between two graphite phases have been obtained by a MC builder 
capable of building polymeric chains of arbitrary geometry in heavily constrained environments,\cite{Macromolecules_47_387} 
based on the quasi-Metropolis scheme of Theodorou and Suter.\cite{Macromolecules_18_1467} 
Then, connectivity altering moves,\cite{PhysRevLett_88_105503} such as concerted 
rotation\cite{MolPhys_78_961,Macromolecules_28_7224} and double bridging\cite{JChemPhys_117_5465} were employed 
in MC equilibration of the initial configurations.
The simulations were carried out in orthorhombic cells with periodic boundary conditions along the $x$ and $y$ directions,
while the system along the $z$ direction was considered as finite.
All MD simulations have been conducted using the Large-scale Atomic/Molecular Massively Parallel Simulator 
(LAMMPS),\cite{JCompPhys_117_1} extended with the graphite potential of Steele,\cite{SurfSci_36_317} 
which we have incorporated into the LAMMPS source code. 

\begin{figure}
   \centering
   \includegraphics[width=0.45\textwidth]{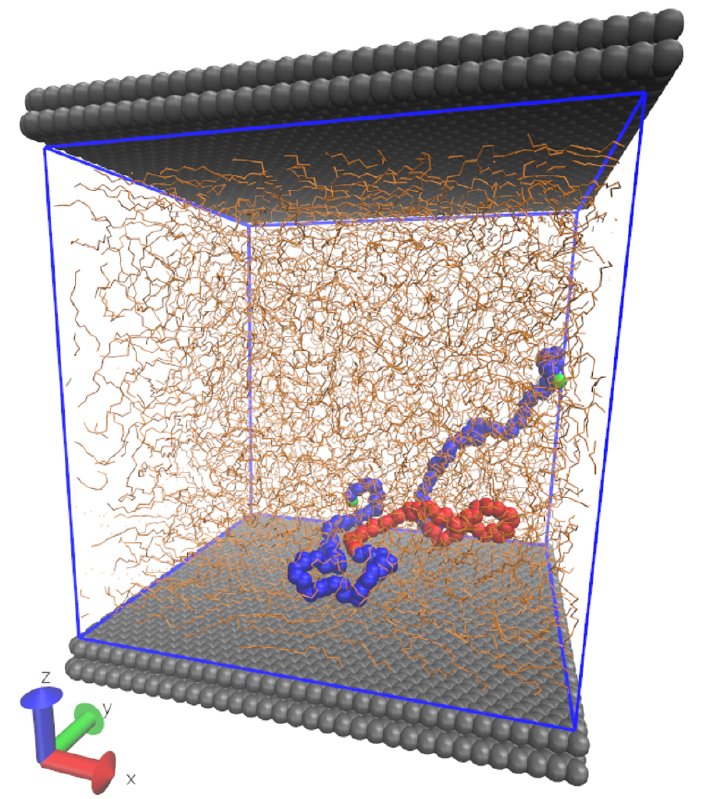}
   \caption{Schematic illustration of the end-constrained subchain of 50 \ce{CH2} units (marked in red), whose residence 
   times in the adsorbed and desorbed state, in the course of an MD simulation, were used for estimating adsorption 
   and desorption kinetics via hazard-plot analysis. The chain is depicted as a necklace of the van der Waals 
   spheres of the atoms, with atoms not belonging to the selected subchain shown in blue.
   The ends of the red subchain have been anchored at height $z(0)=z(1)=1.3\;{\rm nm}$.
   The remaining chains in the system are depicted in wireframe representation, for clarity.
   Two indicative layers of graphite atoms lie at the upper and lower $z$ bounds of the simulation box.
   Visualization made by using the VMD software.\cite{JMolGraphics_14_33}}
   \label{fig:md_schematic}
\end{figure}

In our MD simulations, the graphite substrates lie at the $z=0$ and $z=2L_z$ planes (Figure \ref{fig:md_schematic}). 
The simulated polymeric melt consists of $79$ chains of $200$ united atoms (either \ce{CH2} or \ce{CH3}) each. 
All simulations (MC and MD) have been carried out at the temperature of $450 \;{\rm K}$ and lateral 
pressure $p_{zz} = 1 \; {\rm atm}$.
After letting the system relax in the course of an MD simulation, we decide to anchor an internal strand 
(marked in red in Figure \ref{fig:md_schematic}) by
zeroing the forces acting on its ends, having ensured that the velocities of its ends are also zero.
We then monitor the residence time of this strand in the free (all united atoms at $z > \delta$) and adsorbed 
(at least one united atom at $z \le \delta$) states. The length of the runs was in the order of 400 to 500 ns, by 
employing the RESPA multiple-timestep integrator\cite{JChemPhys_97_1990} with the outer integration step 
being $4\;{\rm fs}$.

The adhesion tension, i.e., the difference between the surface tension of the clean graphite surface, $\gamma_{\rm s}$, 
and the interfacial tension of the graphite/polymer interface, $\gamma_{\rm fs}$,  
can be computed following the approach developed by Tolman\cite{JChemPhys_16_758} and 
refined by Kirkwood and Buff,\cite{JChemPhys_17_338}
as an integral of the ensemble averaged instantaneous difference between 
the normal and tangential stresses $\sigma_\perp (z)$ and $\sigma_\parallel (z)$:
\begin{equation}
   \left( \gamma_{\rm s} - \gamma_{\rm fs} \right) = - \frac{1}{2} 
   \left\langle
   \int_{0}^{L_z} \left(\sigma_\parallel\left(z\right)- \sigma_\perp\left(z\right) \right) \D z
   \right \rangle
\label{eq_surface_tension_tolman}
\end{equation}
where, in our geometry, $\sigma_\perp (z) = \sigma_{zz} (z)$ and 
$\sigma_\parallel (z) = \left(\sigma_{xx}(z) + \sigma_{yy}(z)\right)/2$.\cite{Macromolecules_24_4295}
Away from an interface, $\sigma_\perp = \sigma_\parallel$ and the integrand vanishes. 
Therefore the nonzero contributions to the integral in  \ref{eq_surface_tension_tolman} come from the 
interfacial region. 
The outer factor of $1/2$ in  \ref{eq_surface_tension_tolman} accounts for the presence of two liquid-solid 
interfaces.
The stress tensor as a function of the distance from the surface can be estimated by summing atomic-level stress 
contributions, $\boldsymbol{\sigma}_i$, of atoms lying in a specific slab along the $z$ direction,
${\boldsymbol \sigma}\left(z\right) = \left(1/V_{\rm slab}\left(z\right)\right) 
\sum_{i \in {\rm slab}} {\boldsymbol \sigma}_i V_i$.\cite{Macromolecules_47_387}
Here, we use a Voronoi tessellation to define atomic volumes, \cite{Chaos_19_041111, PhysRevE_74_021306}
such that summing over all atoms gives the total volume of the system, $V = \sum_i V_i$. Following Mansfield 
and Theodorou\cite{Macromolecules_24_4295} and Theodorou et al.\cite{MacromolTheorySimul_2_191} we define the atomic 
level stress as:
\begin{align}
   \sigma_{i, LM}  = - & \frac{1}{V_i}m_i v_{i,\rm L} v_{i,\rm M} \nonumber \\
   - &\frac{1}{2V_i} \sum_{j \ne i} \left(r_{i,L} - r_{j,L} \right)^{\rm min.im.} \: F_{ij,\:M}^{\rm min.im.} 
   \nonumber \\
   - & \frac{1}{V_i} \sum_{k=1}^{n_{\rm walls}} d_{ik,L} F^{\rm w}_{ik,M} 
   \label{eq_atomic_stress_definition}
\end{align}
where $\mathbf{r}_{i}$ and $\mathbf{r}_{j}$ are the position vectors of atoms $i$ and $j$, 
$\mathbf{v}_i$ the velocity of atom $i$ and $\mathbf{F}_{ij}$ is the 
force exerted on atom $i$ by atom $j$. In general, the force $\mathbf{F}_{ij}$ between two sites $i$ and $j$ is 
defined as $\mathbf{F}_{ij} = - \nabla_{\mathbf{r}_i - \mathbf{r}_j} \mathcal{V}$ where $\mathcal{V}$ is the total 
potential energy of the system and the gradient is taken keeping all intersite separation vectors other than 
$\mathbf{r}_i - \mathbf{r}_j$ constant. 
The superscript ``min.im.'' indicates interatomic distances and forces calculated according to the ``minimum
image convention''.
The indices $L$ and $M$, indicating the three coordinate directions in a Cartesian system, assume the values $x$, $y$ 
and $z$. 
The last term of  \ref{eq_atomic_stress_definition} stems from the interaction of the atoms with the substrate. The
quantity $d_{ik,L}$ multiplying the force exerted by the $k$-th wall, 
$F^{\rm w}_{ik,M}$ on atom $i$, 
takes the values: $d_{ik,x} = x_i$, $d_{ik,y} = y_i$ and 
$d_{ik,z} = z_i - z_{k,{\rm w}}$
with $z_{k,{\rm w}}$
indicating the position of the $k$-th (upper or lower) graphite slab along the $z$-direction. 
Moreover, long-range contributions to the atomic-level stresses of  \ref{eq_atomic_stress_definition} have been
calculated based on eqs 27 and 28 of ref \citenum{JPhysChem_93_7320}.

\section{Results}
\subsection{Self-Consistent Field and Local Volume Fraction}
\begin{figure}
   \centering
   \includegraphics[width=0.45\textwidth]{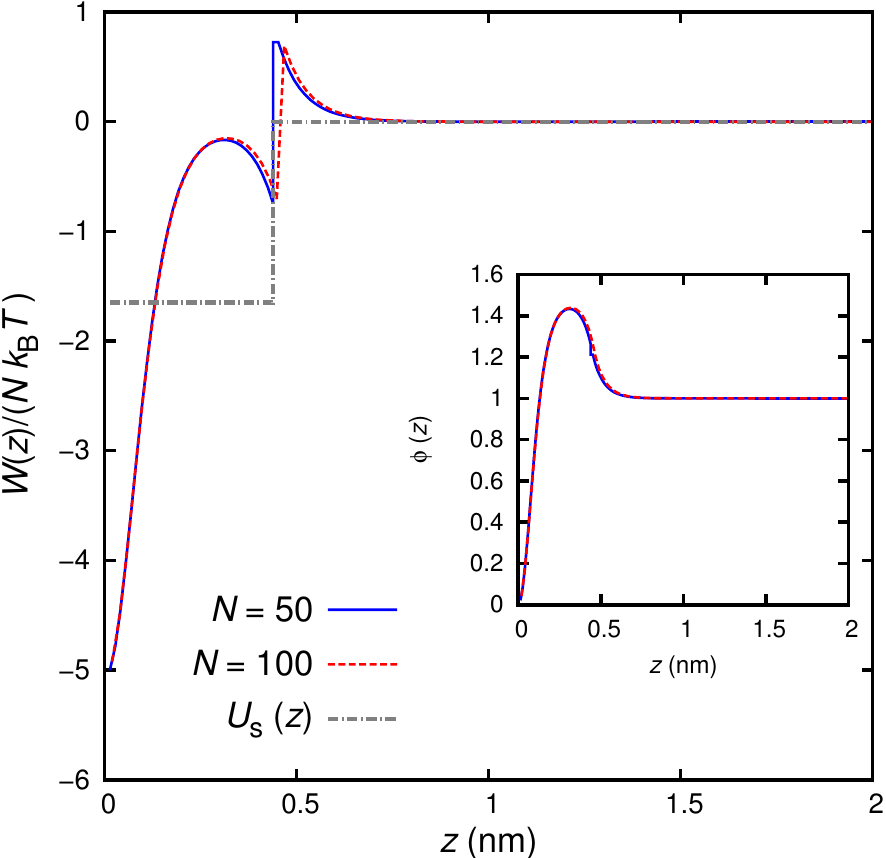}
   \caption{Self consistent field per segment, $W\left( z \right)/\left(N k_{\rm B} T \right)$ for the two different
   chain lengths considered, $N = 50$ and $N = 100$. 
   In the inset to the figure, the local volume fraction is also depicted.}
   \label{fig:fields}
\end{figure}

The SC field obtained from the solution of the Edwards diffusion equation ( \ref{eq_edwards_diffusion_equation})
is presented in Figure \ref{fig:fields}, for two different chain lengths, $N = 50$ and $N = 100$. Along with the field, 
the attractive potential of the solid surface is also shown for comparison. 
The SCF fields exhibit  a smooth behavior due to the fine discretization used for the solution of the SCF problem. 
The square well form of the potential causes a minor numerical instability at the distance where it becomes zero, 
yielding an discontinuous behavior of the SCF 
profiles,at the end of the well.
This is expected and does not affect the following calculations.
The use of a square-well potential is in the spirit of coarse-graining, given the fact that this potential preserves
the same adsorption energy per unit area, and the range of the polymer density fluctuations 
close to the surface as that of the atomistic model.
The use of more detailed potentials is fully compatible with the SCF formulation developed. 
In the inset to Figure \ref{fig:fields} 
the local volume fraction is depicted. The local volume fraction and the structural features obtained from 
the solution of the SCF are the same as the ones by Daoulas et al.\cite{Macromolecules_38_5780}
For more details and a thorough study of several structural features the reader is referred to that work.

\subsection{Adhesion Tension}
Within the maximum term approximation, which leads to  \ref{eq_eff_Hamiltonian_optimum} for the effective Hamiltonian 
at the optimum, the grand potential for the considered system is estimated as:
\begin{align}
   \Omega\left(V,T,\mu\right) = & -k_{\rm B} T \ln{\Xi} \nonumber \\
   = & - \rho_0 \int \D^3 r w^\prime \left({\mathbf r}\right) \phi \left({\mathbf r} \right) \nonumber \\
   & +  \frac{1}{2\kappa_{\rm T}} \int \D^3 r \left[1 - \phi\left({\mathbf r} \right)\right]^2 \nonumber \\
   & + \rho_0 \int \D^3 r U_{\rm s}\left({\mathbf r}\right)  \phi\left({\mathbf r}\right) \nonumber \\
   & - \frac{\rho_0}{N\beta}\int \D^3 r \phi\left({\mathbf r}\right)
   \label{eq_grand_potential}
\end{align}
Eq \ref{eq_grand_potential}, applied to the same volume $V$ of homogeneous polymer in the bulk, which will contain
$n_{\rm bulk}$ chains at chemical potential $\mu$ and temperature $T$, yields:
$\Omega_{\rm bulk} \left(V,T,\mu\right) = - n_{\rm bulk}/\beta = -(\rho_0 V)/(N\beta)$ since 
$w^\prime \left({\mathbf r}\right) = 0$ and $\phi\left({\mathbf r}\right) = 1$ in the bulk.
Introducing the subscripts ``f'' and ``fs'' to distinguish between free polymer surface tension and solid/polymer 
interfacial tension, respectively, and denoting the surface tension of the clean (pure) solid as $\gamma_{\rm s}$,
we define the adhesion tension\cite{Macromolecules_22_4578} as the difference $\gamma_{\rm s} - \gamma_{\rm fs}$.
Now, considering the special case of a polymer melt next to an undeformable solid surface, 
\begin{equation}
   \alpha \left(\gamma_{\rm s} - \gamma_{\rm fs}\right) =  
   \Omega_{\rm bulk}\left(V, T, \mu\right) - \Omega \left(V,T,\mu\right)
\end{equation}
where $\alpha$ is the total interfacial area of contact.
The asymptotic value obtained from the solution of the SCF problem is 
$\left(\gamma_{\rm s} - \gamma_{\rm fs}\right) = 73.6\;{\rm mN/m}$, which is in good agreement with the adhesion tension
estimated by atomistic MD simulations, 
$\left . \left(\gamma_{\rm s} - \gamma_{\rm fs}\right)\right|_{\rm MD} = 70 \pm 10 \;{\rm mN/m}$

To the best of our knowledge, experimental data on interfacial thermodynamic properties of the polyethylene/graphite
system are not readily available. According to the theory of Girifalco and Good\cite{JPhysChem_61_904}
the adhesion tension, can be expressed in terms of the geometric mean of the surface tensions of the liquid and 
the solid as:
\begin{equation}
   \gamma_{\rm s} - \gamma_{\rm fs} \simeq 2 \Phi \left(\gamma_{\rm s} \gamma_{\rm fs}\right)^{1/2} - \gamma_{\rm f}
\end{equation}
The experimentally measured surface tension of PE at $450 \;{\rm K}$ is
$\gamma_{\rm f}^{\rm exp} = 28.1 \;{\rm mN/m}$,\cite{JPhysChem_72_2013,JColloidInterfaceSci_31_153}
and that of graphite is $\gamma_{\rm s}^{\rm exp} = 115 \;{\rm mN/m}$.
\cite{JAmChemSoc_64_2383,JAmChemSoc_68_554,IndEngChem_56_40} 
Based on these values, the experimental estimate of the adhesion tension, taking the interaction parameter
$\Phi = 1$, is $\left(\gamma_{\rm s} - \gamma_{\rm fs}\right)_{\rm exp} = 85.6 \;{\rm mN/m}$.

\subsection{Propagators of End-Constrained Chains}
\begin{figure*}[t]
   \centering
   \includegraphics[width=0.8\textwidth]{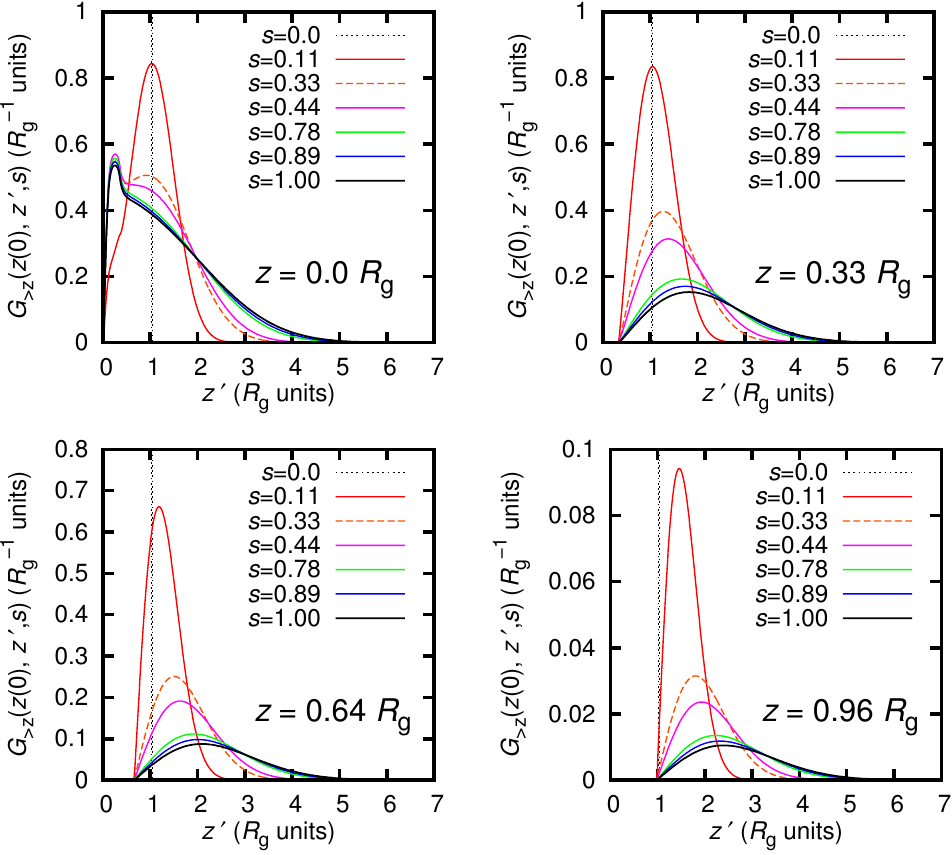}
   \caption{The propagator $G_{>z}\left(z(0), z^\prime, s \right)$ of a chain of length $N=50$ to find its whole
   contour from $z(0)$ to $s$ above the $z$ level, as a function of $z^\prime$. 
   Four different values of the height $z$ have been considered, which are 
   plotted in different subfigures.
   For a given $z$, the distribution becomes broader as $s$ increases.
   The end of the chains has been anchored at height $z(0) = R_{\rm g}$.
   Please note the different scale of the ordinate of the subfigures.}
   \label{fig:propagators_at_rg}
\end{figure*}

In order to calculate the configurational integral of end-constrained chains
based on  \ref{eq_z_final_from_g_integrals}, we have introduced the propagator 
$G_{>z}\left(z(0), z^{\prime}, s \right)$ which is proportional to the probability that a chain
which has started at height $z(0)$ is found at height $z^\prime$, after contour length $s$, 
the whole contour of the chain from 0 to $s$ lying entirely in the region of space $z^{\prime} > z$. 
The propagator is obtained as the solution to a diffusion equation ( \ref{eq_propagator_diffusion}) 
in the presence of the $W(z)$ field and is presented in Figure \ref{fig:propagators_at_rg}.
The chain length considered is $N=50$, while the end of the chains lies at position $z(0) = R_{\rm g}$.

For relatively large values of $z^\prime$, $G_{>z}\left(z(0), z^{\prime}, s \right)$ obeys a Gaussian distribution, 
whose height and breadth increase as $z$ decreases. At low $z^\prime$ values, it drops steeply to zero due to 
the boundary condition $G_{>z}\left(z(0), z^\prime = z, s \right)  = 0$, yielding an overall asymmetry.
For the special case $z=0$ the propagator $G_{>z}\left(z(0), z^{\prime}, s \right)$ resembles a Gaussian distribution 
as long as the chain does not come close to the solid surface. Under the influence of the substrate, the probability
of finding a segment close to it exhibits a local maximum. Thus, for $z=0$ and intermediate values of $s$, the
distribution appears to be bimodal, having one maximum close to $z^\prime = z(0) = R_{\rm g}$ where the $z(0)$ end 
has been anchored, and a second one inside the attractive region close to the substrate.
The distribution becomes broader as the contour length, $s$, increases. This is expected since the chain explores 
more space, moving away from its end.
For $z=0$ the chain segments are likely found inside the adsorption layer, contributing to the local maximum of the
$G_{>z}\left(z(0), z^{\prime}, s \right)$ close to the solid surface. On the contrary, if the dividing surface is 
placed at higher $z > 0$, the propagator is forced to become zero (due to the adsorbing boundary condition 
at $z^\prime = z$), implying also that the probability of finding chain segments close to the surface drops to zero.

\subsection{Configurational Integral of End-Constrained Chains}
\begin{figure}
   \centering
   \includegraphics[width=0.45\textwidth]{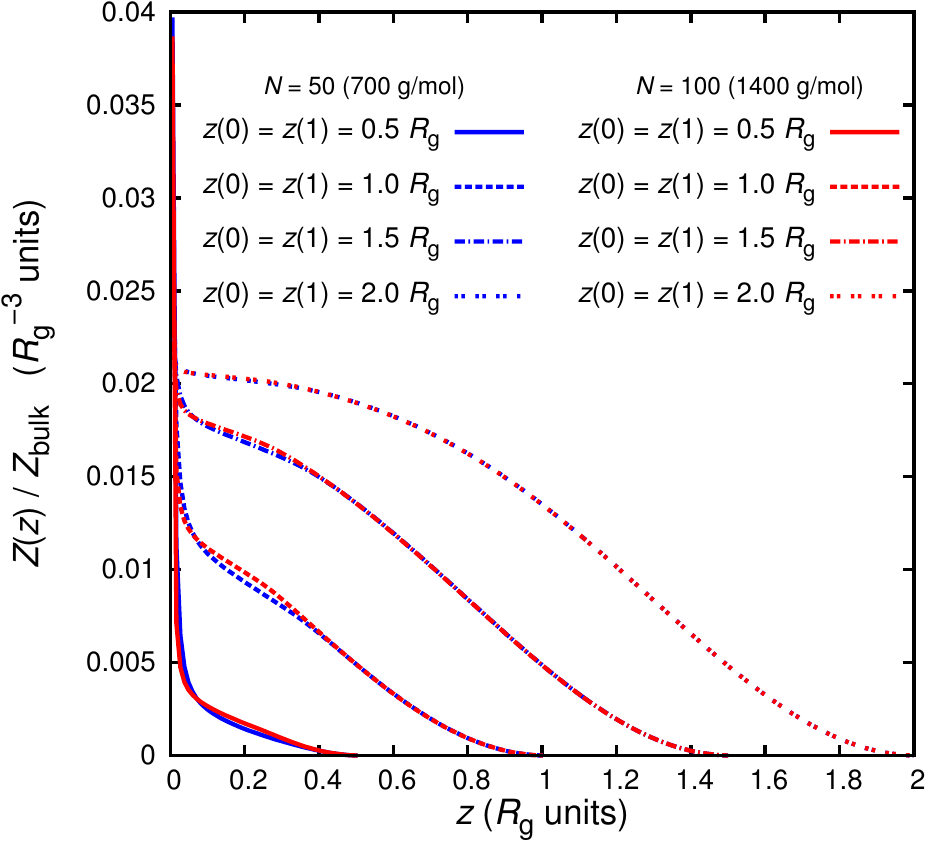}
   \caption{Configurational integral of end-constrained chains, $\mathcal{Z}(z)/\mathcal{Z}_{\rm bulk}$, as a 
   function of distance from the solid surface, for two chain lengths. 
   The ends of the chains are progressively elevated from the solid substrate, starting from 
   $z(0)=z(1)=0.5 R_{\rm g}$ to $z(0) = z(1) = 2.0 R_{\rm g}$.
   The distance between the ends of the strand at an $xy$-level parallel to the surface is $0.5 R_{\rm g}$.}
   \label{fig:conf_integrals}
\end{figure}

Next, we consider the configurational integral $\mathcal{Z}(z)/\mathcal{Z}_{\rm bulk}$ which is proportional to 
the probability that all segments of a chain find themselves at a height greater than $z$ from the substrate.
$\mathcal{Z}(z)/\mathcal{Z}_{\rm bulk}$ is expected to be a decreasing function of $z$, since more polymer chains 
are excluded upon increasing $z$. Moreover it goes smoothly to zero at $z = \min{\left(z(0),z(1)\right)}$, since at least
one end of the chain is found at $z$.
In Figure \ref{fig:conf_integrals}, the dependence of this quantity on the distance of the ends of the strand from the 
solid substrate is presented.
We consider two chains of length $N=50$ and $N=100$ carbon atoms, whose ends are separated by a  
distance of $0.5 R_{\rm g}$ parallel to the surface.
We observe that the profile of $\mathcal{Z}\left(z\right)$ exhibits a spike at $z = 0$. This feature indicates strong
adsorption of the chains onto the solid surface. This effect is more pronounced if the self-consistent field 
$w\left(\mathbf{r}\right)$ consists only of the wall potential, $U_{\rm s}\left(\mathbf{r}\right)$. If the polymer 
field is switched off, chains are extremely strongly attracted by the wall. This results in extremely 
an overwhelmingly high probability of getting adsorbed and monotonically decreasing configurational integrals. 
At the point where both fields (either SCF or wall) approach zero, the results are identical.

For a given distance from the surface, $z$, we observe that the probability of finding a chain for which all segments
lie above this distance increases with the height at which the ends are constrained. This is reasonable, since 
the higher the ends lie from the surface, the more difficult it is for the chain to approach the solid surface, thus
increasing the probability of finding its segments away from it. 
Despite the similarity of the profiles, there exist slight differences between the two different chain lengths 
considered. The differences of the profiles
vanish as we move the ends of the strand higher. This can be attributed to the fact that the range of the attractive
potential is independent of the length of the chains, thus affecting differently the two chain lengths.
At this point, we have to stress that the configurational integrals (as presented in $R_{\rm g}$ units in 
Figure \ref{fig:conf_integrals}) are completely different for the two chain lengths, if considered in the real 
spatial scale of the problem.

\subsection{Equilibrium Probabilities}
\begin{figure*}[t]
   \centering
   \begin{minipage}{.45\textwidth}
      \centering
      \includegraphics{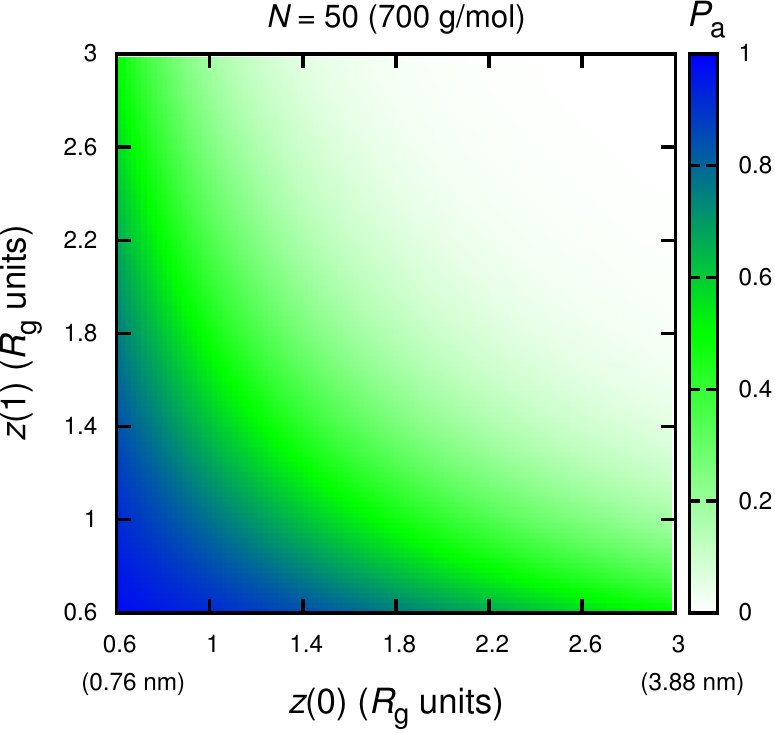} \\* (a)
   \end{minipage}
   \hfill
   \begin{minipage}{.45\textwidth}
      \centering
      \includegraphics{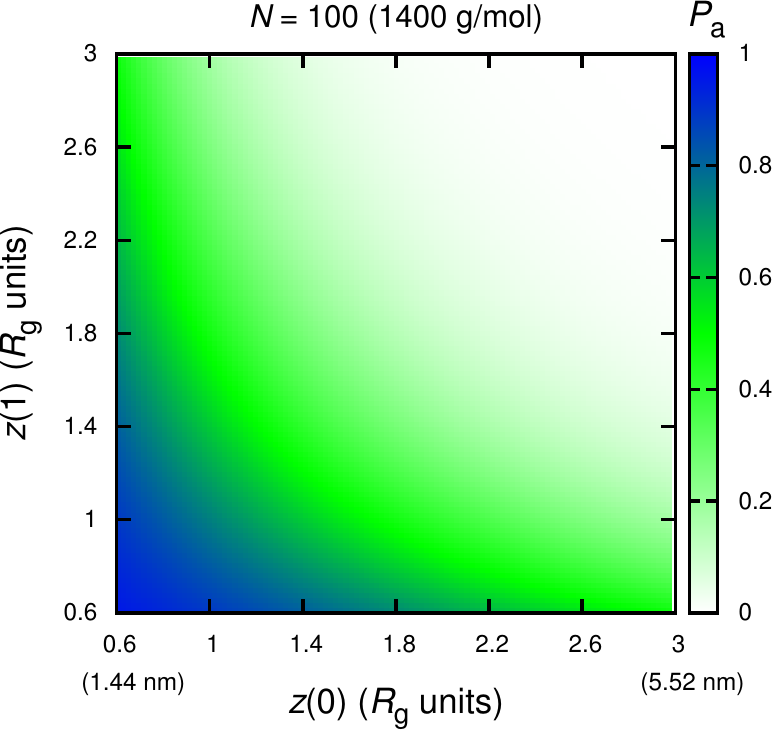} \\* (b)
   \end{minipage}
   
   \caption{Probability of finding a chain in the adsorbed state (at least one segment in the region $z<\delta$)
   as a function of the distance of the ends of the chain from the graphite surface. The end-constrained strand
   consists of either 50 (a) or 100 (b) methylenes. The position of the dividing surface, $\delta$, has been obtained
   as the maximum of the $A(z)$ profiles (see Figure \ref{fig:free_energy_profile_50}).}
   \label{fig:adsorption_probabilities}
\end{figure*}

We now consider the equilibrium probability, $P_{\rm a}$, of finding a chain in the adsorbed state, as this
was defined in terms of the configurational integral ( \ref{eq_equilibrium_probabilities}). 
In Figure \ref{fig:adsorption_probabilities} the contour plots
of the probability to find a chain in the adsorbed state are presented for the two chain lengths considered,
$N=50$, (a) and $N=100$, (b), respectively.
The similarity of the configurational integrals is further manifested in the adsorbed state probability. The contours
are almost identical, given that the spatial scale is expressed in $R_{\rm g}$ units. For comparison, the real distances
(in nm) are also noted in the plots.
For distances of the ends from the surface smaller than or close to $R_{\rm g}$, the probability of finding the 
chain in the adsorbed state (i.e. with at least one of its segments lying below $z = \delta$) is almost unity, due to 
the proximity to the attractive surface. 
For larger distances, the probability of finding the chain in the adsorbed state is reduced, 
exhibiting a roughly hyperbolic dependence on the distances of the ends of the strand from the surface. 
For $z(0), z(1) \simeq 2 R_{\rm g}$ it
is highly unlikely to find an adsorbed chain.
If one of the two ends of the chain remains at a short distance from the surface (e.g. up to $R_{\rm g}$), there 
is always a finite probability of finding the chain in the adsorbed state, irrespectively of the position of its 
other end. This is a consequence of modeling chains as infinitely extensible Gaussian threads. 
Finally, the contour plots are symmetrical about $z(0) = z(1)$. This stems from the fact that the
choice of which end of the chain corresponds to $s=0$ and which corresponds to $s=1$ is arbitrary.

\subsection{Adsorption Potential of Mean Force}
\begin{figure}
   \includegraphics[width=0.45\textwidth]{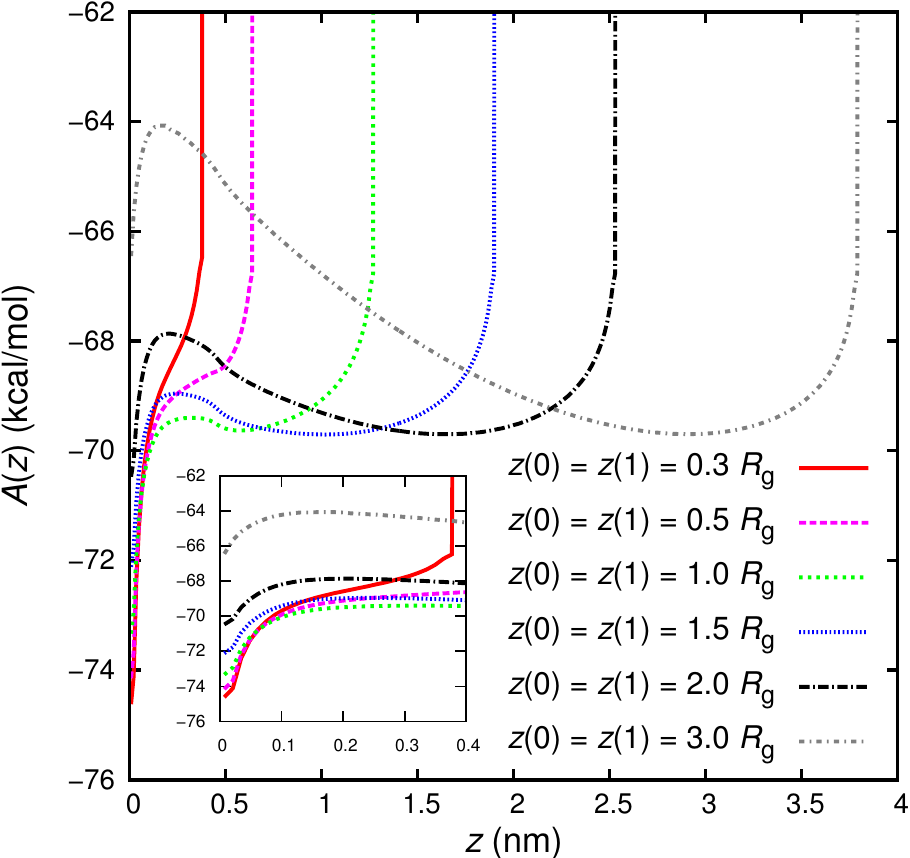}
   \caption{Potential of mean force, $A(z)$, for an end-constrained strand to have its lowest bead at distance $z$ to 
            $z+\D z$ from the solid substrate ( \ref{eq_potential_of_mean_force}). 
            The chain consists of 50 \ce{CH2} groups and its unperturbed radius of gyration is
            $R_{\rm g} = 1.27 \;{\rm nm}$. 
            The normalization constant used in  \ref{eq_potential_of_mean_force} was $C_4 = 1\;{\rm m^4}$.
            In the inset to the figure, the potential of mean force in the region $0 \le z \le 0.4 \;{\rm nm}$ is 
            displayed in order to facilitate the estimation of the depth of the minimum of $A(z)$ close to the
            wall.}
   \label{fig:free_energy_profile_50}
\end{figure}

Based on the configurational integral, $\mathcal{Z}(z)/\mathcal{Z}_{\rm bulk}$, we have defined a free energy per unit 
length, $A(z)$, of a chain to have at least one of its segments below a certain distance, $z$, from the substrate.
To compute $A(z)$ we apply a spline interpolation to $\mathcal{Z}(z)$ in order to obtain a smooth estimate of its 
first derivative, $\D \mathcal{Z}(z)/\D z$, which we then use in order to estimate the free energy profile, $A(z)$, via . 
\ref{eq_potential_of_mean_force}.
The resulting profiles for $N = 50$ and various distances of the ends of the chain from the substrate are depicted 
in Figure \ref{fig:free_energy_profile_50}.
The normalization constant $C_4$ which is needed in  \ref{eq_potential_of_mean_force} has been set to 
$1 \;{\rm m}^4$ in order to cohere with the units of ${\rm m}^3$ in which the configurational integral has been expressed.
The potential of mean force is expected to become infinite for $z \to \min{\left(z(0),z(1) \right)}$, due to the hard
constraint we impose with respect to the height of the ends from the surface.
This behavior is evident in Figure \ref{fig:free_energy_profile_50}, where we have anchored both ends at the same 
height, for simplicity.

The free energy profiles exhibit a well defined minimum extremely close to the solid surface ($z=0$), 
whose depth depends on the distance of the ends of the chain from the surface.
For $z(0)=z(1) < 1.0 R_{\rm g}$ the adsorption of the chains on the surface is spontaneous, the free energy 
with respect to distance being a monotonically increasing function. The only thermodynamically stable state for the chains
is the adsorbed one, in agreement with the adsorption probabilities presented in 
Figure \ref{fig:adsorption_probabilities}. 
The fact that no free state exists for these distances from the wall is reminiscent of the behavior of terminally 
grafted chains in ref \citenum{PhysRevE_87_022604}, which are shown to be exclusively in an adsorbed state when the 
segment-substrate attractive interaction energy exceeds a certain critical value.
For distances of the ends $z(0), z(1) \ge R_{\rm g}$, a second minimum of the
free energy appears, corresponding to a ``free'' or desorbed state of this chain in the bulk. Thus, a barrier separating
the two minima appears. The height of the barrier grows as the ends of the chain are anchored further apart from
the solid surface. 
This is expected, since it becomes progressively more difficult for the chains to expose their segments close to the
surface. Moreover, the position of the minimum shifts to larger distances, indicating that the adsorption becomes
even more difficult.
On the contrary, the position of the barrier seems to remain unaffected upon increasing the 
distance of the ends from the solid surface. This facilitates the clear definition of a dividing surface centered at the
top of the barrier, $\delta \simeq 0.3 \;{\rm nm}$, relative to which adsorption and desorption are considered to take place. 
The maximum of the free energy coincides with the width of the square-well potential (Figure \ref{fig:fields})
and is independent of the length of the chains.
The profile at the top of the barrier can be locally approximated by a polynomial function of high degree (6 or 8), 
allowing us to precisely define its maximum ($\left . \D A(z)/\D z\right|_{z=\delta} = 0$) and obtain an accurate 
estimate of its curvature ($\left . \D^2A(z)/\D z^2 \right|_{z=\delta}$) needed in 
eqs \ref{eq_correction_factor} and \ref{eq_fenergy_curvature} in order to estimate the adsorption and desorption rate 
constants from eqs \ref{eq_adsorption_rate_constant} and \ref{eq_desorption_rate_constant}, respectively.

\subsection{Adsorption and Desorption Rates from SCF/TST}
\begin{figure}
   \centering
   \includegraphics[width=0.45\textwidth]{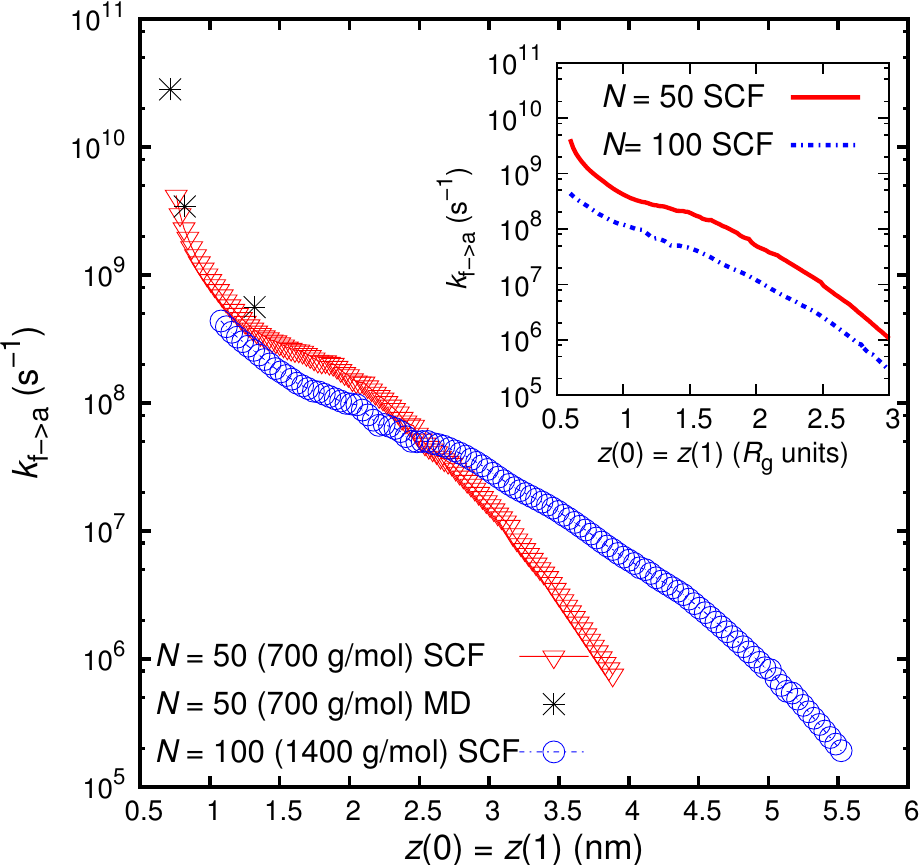}
   \caption{Adsorption rate constant, $k_{{\rm f}\to{\rm a}}$, as a function of the distance of the ends from the
   graphite surface. Both ends are kept at the same distance from the graphite surface, $z(0) = z(1)$ and we consider
   chain lengths of $N = 50$ and $N=100$ methylenes.
   In the inset to the figure the same data are presented with the abscissa in $R_{\rm g}$ units.
   Adsorption rates for an end-constrained \ce{C50} subchain in a \ce{C200} polyethylene melt next to graphite obtained
   by MD simulation using a detailed united-atom model are also shown (compare Figure \ref{fig:hazard_plot}).}
   \label{fig:adsorption_rate_MD_comparison}
\end{figure}

The last and most crucial step in our study is the estimation of adsorption and desorption rate constants for an
end-constrained strand. The identification of the free energy barrier height (Figure \ref{fig:free_energy_profile_50}) 
and the estimation of its curvature allows us to apply dynamically-corrected transition state theory to 
extract the rate constants for a strand to overcome the barrier of free energy, in order to either adsorb or desorb.
In the following, we will use $\zeta_{\rm strand} = \left(m_{\rm strand} / m_{{\rm CH}_2} \right) \zeta_0$ as the 
friction coefficient of the strand, invoked in eqs \ref{eq_adsorption_rate_constant} 
and \ref{eq_desorption_rate_constant}. 
At first, in Figure \ref{fig:adsorption_rate_MD_comparison} we restrict ourselves to equidistant ends of the strand
from the solid surface. 
The rate constants we present are independent of the distance between the projections of the ends on the 
$xy$ plane, due to the factorizability of the propagators ( \ref{eq_z_from_g_integrals}) that leads to cancellation 
between numerator and denominator of the last terms in eqs \ref{eq_adsorption_rate_constant} and 
\ref{eq_desorption_rate_constant}.
Based on  \ref{eq_adsorption_rate_constant} we estimate the adsorption rate constant of 
chains of length 50 and 100. The results are presented together with the estimates of adsorption rate constant
obtained by hazard-plot analysis\cite{JPhysChemB_112_10619} of MD trajectories for the same system (see next 
subsection).\cite{atomisticMD} 

The adsorption rate, as presented in Figure \ref{fig:adsorption_rate_MD_comparison}, exhibits an extremely strong 
dependence on the distance of the strand's ends from the solid surface. The nearly exponential decay of the rate
constant with distance of the constrained ends from the surface can be observed for both chain lengths 
under consideration. 
In the inset to Figure \ref{fig:adsorption_rate_MD_comparison}, the same data are presented with the abscissa 
measured in $R_{\rm g}$ units. 
For both chain lengths, the estimation of the rates can be accomplished down to $0.5 R_{\rm g}$; below this lower 
limit adsorption is spontaneous, leading to infinitely high adsorption rates.
At intermediate distances, a nearly exponential decay of the adsorption rate constant takes place, for both chain lengths
considered. At even larger distances, a downward departure from the 
exponential trend can be observed. The adsorption rate should go to zero when the triangle formed by the end-to-end
distance and two fully extended substrands of total length equal to the chain length, forming equal angles with 
the surface normal, barely reaches the surface. 
The infinite extensibility of the Gaussian model may account for the gentler drop of the adsorption rate constant, as
observed in Figure \ref{fig:adsorption_rate_MD_comparison}. 
The adsorption rate constants depend also on the molar mass of the strands.
At small distances from the surface, shorter, thus more mobile, chains can approach the surface more easily. However, 
at larger distances, where stretching of the chain should take place to reach the surface, longer chains with more
segments adsorb faster. 
The time scale at which adsorption takes place cannot be fully accessed by MD simulations, especially for the longer 
chains. However, where results are available from direct measurement of adsorption rates, the agreement between the
SCF/TST approach and the MD simulations is favorable.

Rate constants for adsorption/desorption from the melt are difficult to measure directly. 
Dietsche et al.\cite{AIChEJ_41_1266} have measured the displacement of deuterated alkanes by hydrogenous alkanes in an 
\ce{C16} alkane liquid flowing past a surface (\ce{ZnSe}, \ce{Cr}, \ce{Cu}) with attenuated total reflectance Fourier 
transform infrared spectroscopy (ATR/FTIR).
The ATR/FTIR experiments revealed a surprising interaction between the weakly adsorbing oligomeric polyolefins and
the surface, which manifests itself in slow exchange between the surface and the bulk.
The picture which seems to emerge is an entropically-driven flow-rate- (or stress-) 
dependent adsorption equilibrium which retards the exchange between the surface and bulk. 
In order to reproduce the experimental findings with finite-element calculations, these authors have assumed a 
diffusivity of $3.1 \times 10^{-10} \;{\rm m}^2\;{\rm s}^{-1}$. If we imagine a strand diffusing in the interfacial region with
this diffusivity, the timescale of the diffusion of a strand over a lenght commensurate with its radius of gyration is
$\sim 10^{-8} \;{\rm s}$, which is relevant to the results obtained by the SCF/TST approach.

\begin{figure*}
   \centering
   \begin{minipage}{.45\textwidth}
      \centering
      \includegraphics{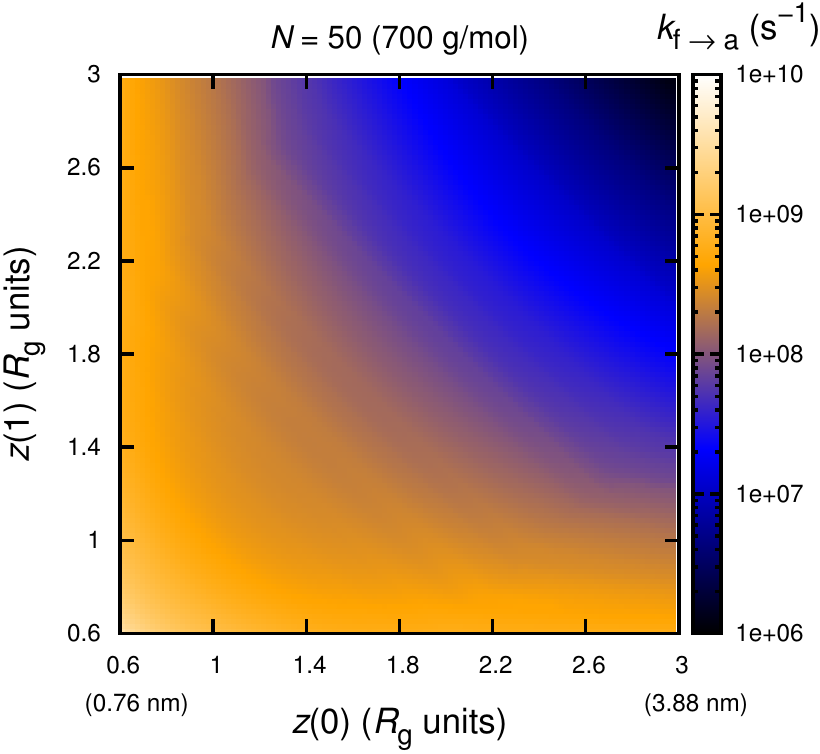} \\* (a)
   \end{minipage}
   \hfill
   \begin{minipage}{.45\textwidth}
      \centering
      \includegraphics{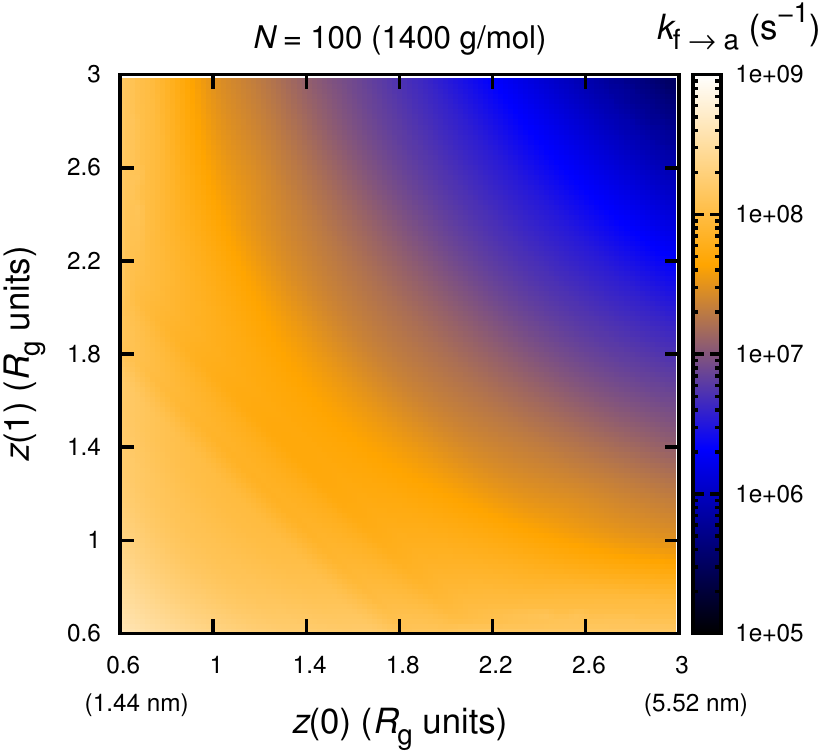} \\* (b)
   \end{minipage}
   
   \caption{Adsorption rate constant, $k_{{\rm f}\to {\rm a}}$, as a function of the distance of the ends of 
   the chain from the graphite suface.
   The chain consists of either $50$ (a), or $100$ (b) methylenes. The radii of gyration are 
   $R_{\rm g}(50) = 1.27 \;{\rm nm}$ and $R_{\rm g}(100) = 1.80 \;{\rm nm}$, respectively.}
   \label{fig:adsorption_rates}
\end{figure*}

The next step, Figure \ref{fig:adsorption_rates}, is to allow the distances of both ends of the chain from the 
substrate to vary independently.
The adsorption rate contour plot, Figure \ref{fig:adsorption_rates}, is symmetrical about $z(0) = z(1)$,
since the choice of which end of the chain corresponds to $s=0$ and which corresponds to $s=1$ is arbitrary.
It is evident that the rate constants extend over a wide range of timescales 
(from $k_{{\rm f} \to {\rm a}} \sim 10^{10}\;{\rm s}^{-1}$ at $1 \; {\rm nm}$ 
to $k_{{\rm f} \to {\rm a}} \sim 10^{5}\;{\rm s}^{-1}$ at $5 \; {\rm nm}$). 
The adsorption rate for both chain lengths (Figure \ref{fig:adsorption_rates} (a) and (b)) drops quickly when moving
away from the surface.
It should be noted that the adsorption rate drops almost by one order of magnitude upon elevating the ends of the 
strand by 1 nm from the graphite surface. 
However, if one of the ends of the chain finds itself close to the solid substrate, the
adsorption rate constant remains relatively high, irrespectively of the height of the other end.
Overall, at the same values of $z(0)/R_{\rm g}$ and $z(1)/R_{\rm g}$, longer chains 
(Figure \ref{fig:adsorption_rates} (b)) adsorb more difficultly on the solid surface, than shorter
chains do (Figure \ref{fig:adsorption_rates} (a)).
This may be attributed to the increased friction they experience due to interactions with their environment, 
resulting in reduced mobility.

\begin{figure*}
   \centering
   \begin{minipage}{.45\textwidth}
      \centering
      \includegraphics{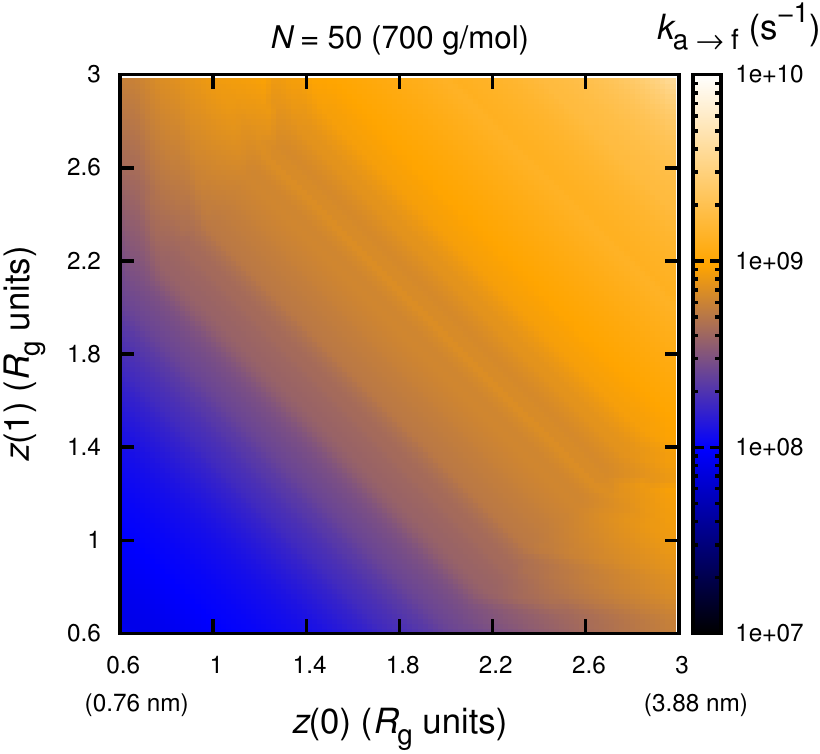} \\* (a)
   \end{minipage}%
   \hfill
   \begin{minipage}{.45\textwidth}
      \centering
      \includegraphics{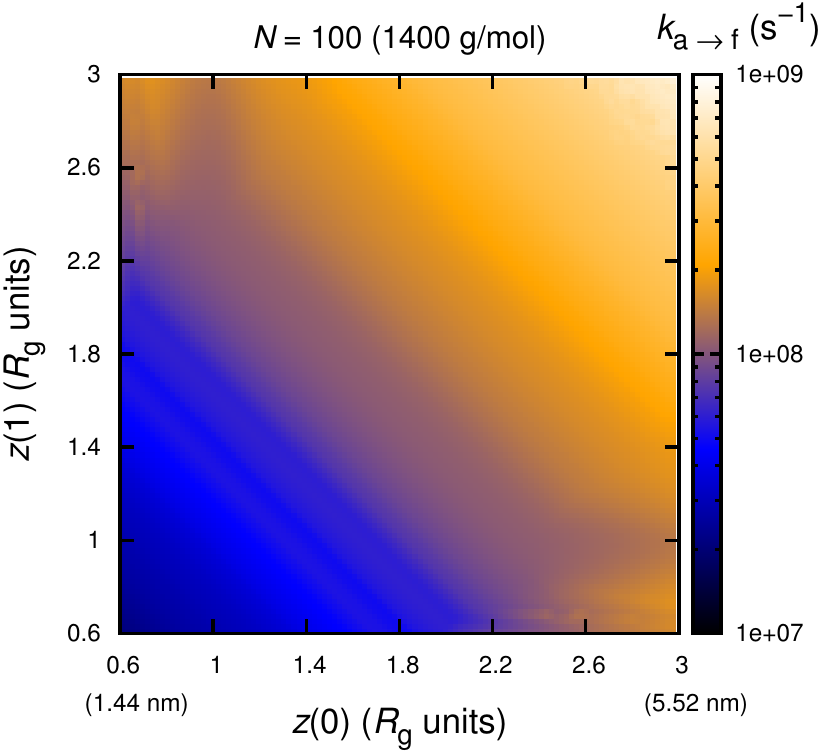} \\* (b)
   \end{minipage}
   
   \caption{Desorption rate constant, $k_{{\rm a}\to {\rm f}}$, as a function of the distance of the ends of 
   the chain from the graphite suface.
   The chain consists of either $50$ (a), or $100$ methylenes. The radii of gyration are 
   $R_{\rm g}(50) = 1.27 \;{\rm nm}$ and $R_{\rm g}(100) = 1.80 \;{\rm nm}$, respectively.}
   \label{fig:desorption_rates}
\end{figure*}

The dependence of the desorption rate on the distance of the ends from the substrate is presented in Figure 
\ref{fig:desorption_rates}. As expected, desorption rates are increasing functions of the distance of the ends of 
the chain from the surface and exhibit a symmetry about $z(0) = z(1)$, since the assignment
of the chain ends to $s=0$ and $s=1$ is arbitrary. 
Close to the surface the desorption rates are lower than the corresponding adsorption 
rates. As one moves further away from the substrate, however, the balance of the two rate constants radically changes. 
Shorter chains exhibit the tendency to desorb faster, under the same conditions.
For both adsorption and desorption rates, the detailed balance condition holds. This means that the probability of
being free multiplied by the adsorption rate constant equals the probability of being adsorbed multiplied by the 
desorption rate constant, for all cases considered.

\subsection{Adsorption Rates from MD}
In order to evaluate the rate constants for the adsorption/desorption events, we invoke a hazard plot 
analysis\cite{JChemPhys_69_1010, Macromolecules_13_526, JPhysChemB_112_10619} of the atomistic MD trajectory.
Hazard plot analysis is based on the evaluation of the cumulative hazard.
The hazard rate, $h(t)$, is defined such that $h(t)\D t$ is the probability that a system which has survived a 
time $t$ in a certain state since its last transition, will undergo a transition at the time between $t$ and $\D t$. 
The cumulative hazard is defined as $H(t) = \int_0^t h\left(t^\prime\right)\D t^\prime$.
By assuming a Poisson process consisting of elementary transitions with first order kinetics from one state to 
the other, the Poisson rate can be extracted as the slope of a plot of the cumulative hazard versus the residence
time. Thus, the adsorption rate constant can be obtained from the hazard plot of the residence time of the system
in the free state, and vice versa.
The dividing surface of the adsorbed and the free state, is placed at $\delta = 0.3\;{\rm nm}$, as estimated from the
SCF calculations.

\begin{figure}
   \centering
   \includegraphics[width=0.45\textwidth]{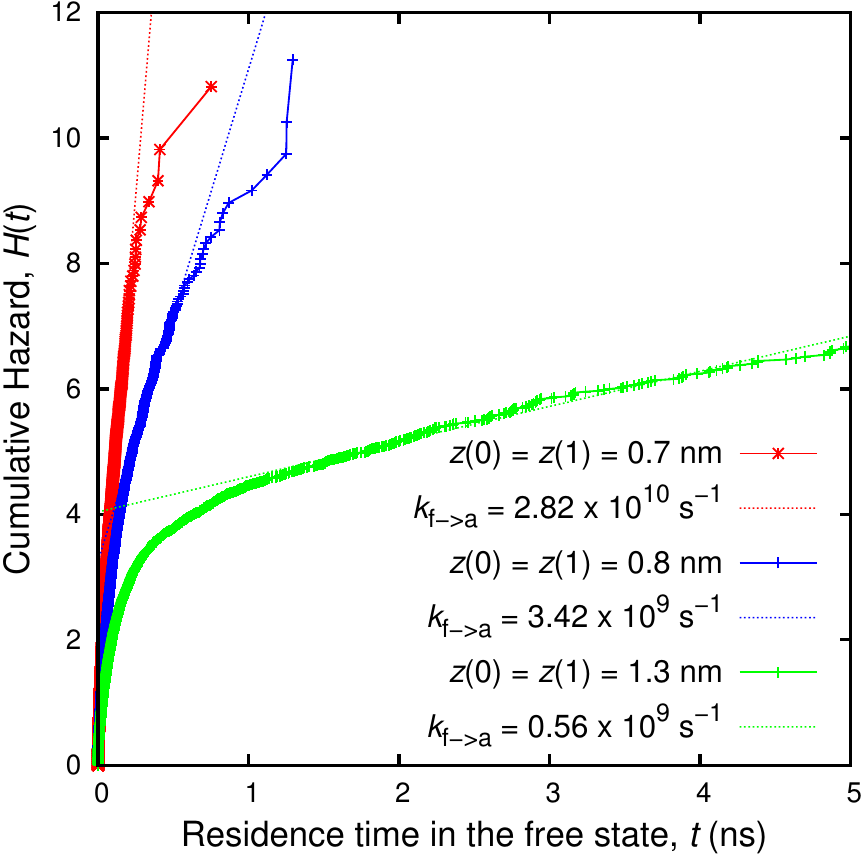}
   \caption{Hazard plot of the time between transitions from the free to the adsorbed state, as calculated 
   atomistically with molecular dynamics. The height at which the 
   ends of the chain are anchored is varied. Linear fits are also included which provide an estimate of the slope, i.e.
   the adsorption rate constant.}
   \label{fig:hazard_plot}
\end{figure}

In Figure \ref{fig:hazard_plot} we present the adsorption hazard plot for an end-constrained intra-chain strand 
of $50$ \ce{CH2} united atoms (such as the red one in Figure \ref{fig:md_schematic}), whose ends have been anchored 
at $z(0) = z(1) = 0.7, 0.8 $ and $1.3\;{\rm nm}$. 
$H(t)$ becomes linear at long times. At short times it exhibits some nonlinearity, especially when $z(0)=z(1)$ is 
large, betraying fast recrossing events at the dividing surface.  
The slope of the curves corresponds to the rate constants
for this strand to make a transition from the free to the adsorbed state. 
It is evident that the time a strand spends in the free state before adsorbing onto the solid substrate grows
quasi exponentially with the distance of the ends from the surface. With our computational resources, a study for 
anchoring distances larger than 1.3 nm was not feasible, due to the maximum time that could be accessed with an
MD simulation. Even for smaller distances, there is a significant error related to fitting the long time part of
curves to a straight line. This is due to the fact that at short time-scales a lot of short-lived events are 
monitored because of the recrossing of the dividing surface.  
The curves depicted in Figure \ref{fig:hazard_plot} provided us with the estimates of adsorption rate constants 
included in Figure \ref{fig:adsorption_rate_MD_comparison}, where the comparison against the SCF/TST estimates 
was made.

\section{Summary and Conclusions}
In this paper we have outlined a strategy that combines field theoretic modeling with
transition state theory to compute rate constants for adsorption and desorption at a melt/solid 
interface of a (sub)chain whose ends are constrained in space, e.g. through crosslinks or entanglements.  
At first, the problem of determining the equilibrium properties of a polymer melt of specific chemical constitution,
adsorbed on a certain solid substrate through a continuum SCF approach was considered, following the 
work of Daoulas et al.\cite{Macromolecules_38_7134} 
The Gaussian chain model reproducing polyethylene chains has been used. As far as the interactions with the substrate
are concerned, these were taken into account via a square well potential whose parameters were matched to 
reproduce the graphite-polyethylene interaction.
The interfacial structure and thermodynamics were obtained from the solution of the SCF problem. 
An estimate of the adhesion tension is provided, for the first time to the best of our knowledge, which has been 
found to be in reasonable agreement with that computed by a realistic united-atom simulation and with available 
experimental evidence.

Based on the solution of the SCF problem, a conveniently defined propagation function allowed us to estimate the 
configurational integral of end-constrained chains with a given distance of closest approach to the
solid substrate. From the configurational integral we computed the Helmholtz energy (potential of mean force) of 
the end-constrained chain as a function of our chosen one-dimensional order parameter (the distance of closest 
approach to the solid surface).
Chains having their ends constrained close to the substrate exhibit a monotonically increasing free energy when 
moving away from it, with a well-defined minimum close to it. Thus, spontaneous adsorption of these chains 
takes place. Chains whose ends are constrained at larger distances from the substrate exhibit a free energy
profile with two well-defined minima (states). The free energy barrier in between the two minima serves for defining 
a dividing surface which discriminates between ``adsorbed'' (close to the wall) and ``free'' (far from the wall) states.
The distance of the barrier from the substrate surface is very close to the width of the square well potential used 
for representing the graphite/polymer segment interaction and is independent of the chain length.

Dynamically corrected transition state theory, in the form of Kramers theory in the high friction limit, applied to 
the free energy profile, provides a reasonable way of describing the dynamics of adsorption/desorption of the 
end-constrained chains. 
The friction factor needed for calculating rates is connected to the monomeric friction coeffecient, which has been
extracted from atomistic MD simulations in the Rouse regime.\cite{Macromolecules_31_7934}
The adsorption/desorption rate constants are found to depend strongly on the distance of the ends of the strands
from the solid surface. The length scale of adsorption is limited to a few nanometers for chains composed of 
50 to 100 methylene units. Despite the locality of the effect, a decay by four orders or magnitude (or more) of the 
adsorption rate constant is observed when moving from to 1 to 5 nm from the graphite surface. 
The rate constant for adsorption decays practically exponentially with the distance of the chain ends from the solid 
surface over an intermediate range of distances between the longest distance where spontaneous adsorption takes place 
and three times the unperturbed mean square radius of gyration.
For larger distances of the ends from the surface the rate constant for adsorption drops more precipitously.  
The rate constant for desorption increases with distance between the chain ends and the solid substrate.  
It is generally less sensitive to distance than the adsorption rate constant.  

To the best of our knowledge, this the first time that the solution
of the polymer SCF problem next to a solid surface has been used to provide insight into the dynamics of the system.
This work suggests that SCF, coupled with transition-state theory, can give good results for the rate of 
adsorption/desorption between polymers in a melt and a surface.
SCF theory is less computationally demanding than atomistic simulations, and thus SCF can be used to examine regimes 
not practically accessible to atomistic simulations, such as the regime of low adsorption rates. Furthermore, 
the parameters in SCF have been directly connected to atomistic parameters, suggesting that SCF may provide a more 
faithful description of the problem than alternative, coarse-grained simulation methods.
In the regime where molecular dynamics simulation of adsorption rates is practical (short chains, very close to the 
surface), our results for the adsorption rate are in good agreement with the results from molecular dynamics 
simulation. Finally, a reasonable agreement between the melt adhesion tension computed from SCF and that measured in 
atomistic simulations and in experiment, was found.

In the ongoing debate of interfacial dynamics, our results suggest that the crosslinks or entanglements that are 
nearest to the surface would be the controlling factors influencing the surface dynamics.
Some of these links may be permanently attached to the surface, while other may adsorb/desorb with a rate constant. 
If the magnitude of the adsorption rate constant is larger than the time needed to entagle/disentagle, the surface 
may enhance the entanglement density of the polymer, thus contributing to its mechanical properties. If the two 
surfaces come closer, in a way that segments lying in the middle can adsorb on both of them, bridges of adsorbed may 
be formed, slowing down the dynamics of the polymer in the slit.

\onecolumn
\section*{Appendix A: Grand Partition Function of an Ensemble of Noninteracting Chains Subject to a Field}
We consider a random fluctuating field $w\left({\mathbf r}\right)$ and form the functional integral:
\begin{equation}
   \int \mathcal{D} \left[\beta w\left({\mathbf r} \right) \right] \exp \left\{ 
   \int \D^3 r \left[ -\imun \left(\hat{\phi}\left({\mathbf r}\right) - 1\right) \beta \rho_0 w\left({\mathbf r}\right)   
   - \frac{\kappa_{\rm T} k_{\rm B} T}{2}\left( \beta \rho_0 w\left({\mathbf r}\right)\right)^2  
   \right] \right\}
   \tag{A.1} 
\end{equation}
with $\imun$ being the imaginary unit, $\beta = 1/\left( k_{\rm B}T\right)$ and the volume fraction operator
$\hat{\phi}\left(\mathbf{r}\right)$ given by  \ref{eq_phihat_definition}. 
We can rearrange the internal integral and write the functional integral as:
\begin{equation}
   \int \mathcal{D} \left[\beta w\left({\mathbf r}\right) \right]
   \exp{\left\{- \mathlarger \int \D^3 r \left[ \left(\frac{\kappa_{\rm T}}{2 k_{\rm B}T}\right)^{1/2} 
   \rho_0 w\left({\mathbf r}\right)
   + \imun \frac{\hat{\phi}\left({\mathbf r} \right) - 1}{\left(2 \kappa_{\rm T} k_{\rm B} T \right)^{1/2}}
   \right]^2 \right\}} 
   \exp{\left[ -\mathlarger \int \frac{\left(\hat{\phi}\left({\mathbf r}\right) - 1 \right)^2}
   {2 \kappa_{\rm T} k_{\rm B} T} \D^3 r
   \right]}
   \tag{A.2}
\end{equation}
The functional integral over $w\left({\mathbf r}\right)$ in the last expression is Gaussian and should yield a number 
$C^{-1}$ as a result, which is dependent upon $\rho_0$, $\kappa_{\rm T}$, and $T$. Thus, we can write
\begin{align}
   \exp{\left[-\mathlarger\int \frac{\left(\hat{\phi}\left( {\mathbf r}\right) - 1\right)^2}
   {2 \kappa_{\rm T} k_{\rm B}T }\D^3 r \right]} = 
   C \int \mathcal{D} \left[\beta w\left({\mathbf r}\right) \right]
   \exp\left\{\mathlarger\int \D^3 r \left[ -\imun\left(\hat{\phi}\left({\mathbf r}\right)-1 \right)\beta 
   \rho_0 w\left({\mathbf r}\right) 
   - \frac{\kappa_{\rm T} k_{\rm B}T}{2}\left(\beta \rho_0 w\left({\mathbf r}\right)\right)^2
   \right] \right\}
   \tag{A.3} \label{eq_appendixA_3}
\end{align}
Substituting the left-hand side of  \ref{eq_appendixA_3} with its right-hand side within 
 \ref{eq_grand_partition_function} we obtain:
\begin{align}
   \Xi = & \sum_{n=0}^{\infty}\frac{1}{n!}\exp{\left(\frac{\mu N n}{k_{\rm B}T} \right)} \tilde{N}^n C
   \int \mathcal{D}\left[\beta w\left({\mathbf r}\right) \right] 
   \int \prod_{\alpha = 1}^{n} \mathcal{D} {\mathbf r}_{\alpha}
   \left( \bullet \right) \mathcal{P} \left[{\mathbf r}_\alpha \left( \bullet \right) \right]
   \nonumber \\
   & \exp\left\{\int \D^3 r \left[- \imun \beta \rho_0 w\left({\mathbf r}\right)
   \left(\hat{\phi}\left({\mathbf r} \right) -1 \right) - \frac{\rho_0}{k_{\rm B}T}
   U_{\rm s}\left({\mathbf r} \right) \hat{\phi}\left({\mathbf r}\right)
   - \frac{\kappa_{\rm T}k_{\rm B}T}{2} \left( \beta \rho_0 w\left({\mathbf r}\right)\right)^2
   \right]  \right\}
   \tag{A.4} \label{eq_appendixA_4}
\end{align}
and, recalling  \ref{eq_phihat_definition},
\begin{align}
   \Xi  = & \sum_{n=0}^{\infty} \frac{1}{n!}\exp{\left(\frac{\mu N n}{k_{\rm B}T} \right)} \tilde{N}^n C
   \int \mathcal{D}\left[\beta w\left({\mathbf r}\right)\right] \nonumber \\
   &\prod_{\alpha = 1}^{n} \left\{ \int \mathcal{D}{\mathbf r}_{\alpha}\left(\bullet \right) 
   \mathcal{P}\left[{\mathbf r}_\alpha \left(\bullet \right) \right]
   \exp{\left[ - \int \D^3 r \int_0^1 \D s \: \delta\left({\mathbf r}-{\mathbf r}_\alpha \left(s\right) \right) 
   \left(\beta\imun N w\left({\mathbf r}\right) + \beta N U_{\rm s}\left({\mathbf r}\right)  \right)
   \right]} \right\} \nonumber \\
   & \exp{\left\{ \int \D^3 r \left[ \imun \beta \rho_0 w\left({\mathbf r}\right) 
   - \frac{\kappa_{\rm T} k_{\rm B} T}{2} \left(\beta \rho_0 w\left({\mathbf r}\right) \right)^2
   \right] \right\}}
   \tag{A.5} \label{eq_appendixA_5}
\end{align}
In  \ref{eq_appendixA_5}, the integration over ${\mathbf r}_\alpha \left(\bullet\right)$ is reduced to a partition 
function of $n$ independent Gaussian chains, the segments of each chain interacting with the field
$\imun w\left({\mathbf r}\right) + U_{\rm s}\left({\mathbf r}\right)$. Introduction of the fluctuating field 
$w\left({\mathbf r}\right)$ has served to decouple the chains and reduce the problem of $n$ interacting chains to  a
problem of $n$ single chains interacting with an effective field in the calculation of the grand partition function.

Following Edwards\cite{Doi_Edwards_TheoryOfPolymerDynamics}, we introduce the notation 
$Q \left[\imun w + U_{\rm s} \right]$ to indicate the partition function of a single chain subject to the field 
$\imun w + U_{\rm s}$, acting on its segments, \emph{relative} to the partition function of a field-free chain.
Then,
\begin{align}
   & \int \mathcal{D} {\mathbf r}_\alpha \left(\bullet\right) 
   \mathcal{P} \left[ {\mathbf r}_\alpha \left(\bullet \right) \right]
   \exp{\left[- \int \D^3 r \int_{0}^{1} \D s \: \delta\left({\mathbf r} - {\mathbf r}_\alpha\left(s \right) \right) 
   \left(\beta \imun N w\left({\mathbf r} \right) + \beta N U_{\rm s} \left({\mathbf r} \right) \right) \right]}
   \nonumber \\
   = & \left\{\int \mathcal{D} {\mathbf r}_\alpha \left(\bullet\right) 
   \mathcal{P}\left[{\mathbf r}_\alpha \left(\bullet\right) \right] \right\}
   Q\left[\imun w + U_{\rm s}\right] \nonumber \\
   = & Z_{\rm free} Q \left[\imun w + U_{\rm s}\right]
   \tag{A.6} \label{eq_appendixA_6}
\end{align}
where $Z_{\rm free} = \int \mathcal{D} {\mathbf r}_\alpha \left(\bullet\right) \mathcal{P}\left[{\mathbf r}_\alpha 
\left(\bullet\right) \right]$ is the partition function of a free chain. With this notation, 
\begin{align}
   \Xi = C\sum_{n = 0}^\infty \frac{1}{n!} & \exp{\left(\frac{\mu N n}{k_{\rm B}T} \right)} 
   \left(\tilde{N} Z_{\rm free} \right)^n \int \mathcal{D}\left[\beta w \right] 
   \left( Q\left[\imun w + U_{\rm s}\right]\right)^n \nonumber \\
   & \exp{\left\{ \int \D^3 r \left[ \imun \beta \rho_0 w\left({\mathbf r}\right)
   - \frac{\kappa_{\rm T} k_{\rm B} T}{2}\left(\beta \rho_0 w\left({\mathbf r}\right)\right)^2
    \right] \right\}}
    \tag{A.7} \label{eq_appendixA_7}
\end{align}
or
\begin{align}
   \Xi = C\int \mathcal{D}\left[\beta w\right] & \exp{\left\{ 
   - \int \D^3 r \left[- \imun \beta \rho_0 w\left({\mathbf r}\right) + \frac{\kappa_{\rm T}k_{\rm B}T}{2} 
   \left(\beta \rho_0 w\left({\mathbf r}\right)\right)^2 \right] \right\}} \times \nonumber \\
   & \sum_{n = 0}^\infty \frac{1}{n!}\exp{\left(\frac{\mu N n}{k_{\rm B}T} \right)}
   \left(\tilde{N} Z_{\rm free} \right)^n
   \left(Q\left[\imun w + U_{\rm s} \right] \right)^n
   \tag{A.8} \label{eq_appendixA_8}
\end{align}
Now, the summation over different $n$ can be performed, yielding
\begin{align}
   & \sum_{n=0}^\infty \frac{1}{n!} \exp{\left(\frac{\mu N n}{k_{\rm B} T} \right)}
   \left(\tilde{N} Z_{\rm free}\right)^n \left(Q\left[\imun w + U_{\rm s}\right]\right)^n
   \nonumber \\
   & = \exp{\left\{\exp{\left(\frac{\mu N}{k_{\rm B}T}\right)} \tilde{N} Z_{\rm free} 
   Q\left[\imun w + U_{\rm s}\right] \right\}}
   \tag{A.9} \label{eq_appendixA_9}
\end{align}
which leads to the expression of  \ref{eq_grand_partition_function_effective_hamiltonian} for the grand partition 
function.

\section*{Appendix B: Saddle-point Approximation}
We replace the functional integral appearing in  \ref{eq_grand_partition_function_effective_hamiltonian} 
of the main text for $\Xi$ with its dominant term, obtained by setting the functional derivative of $H$ to zero, 
${\delta H}/{\delta w} = 0$:
\begin{equation}
   \frac{\delta H}{\delta w} = - \imun \rho_0 + \kappa_{\rm T} \rho_0^2 w\left({\mathbf r}\right) -
   \frac{1}{\beta} \exp{\left(\frac{\mu N}{k_{\rm B}T}\right)} \tilde{N} Z_{\rm free} 
   \frac{\delta Q}{\delta w}
   \tag{B.1} \label{eq_appendixB_1}
\end{equation}
Now, by definition,\cite{Doi_Edwards_TheoryOfPolymerDynamics} 
 \ref{eq_appendixA_6} of Appendix A can be rewritten as:
\begin{equation}
   Q\left[ \imun w + U_{\rm s}\right] = \frac{\mathlarger \int \mathcal{D}{\mathbf r}_\alpha \left(\bullet\right)
   \mathcal{P}\left[{\mathbf r}_\alpha \left(\bullet\right)\right]
   \exp{\left\{-\mathlarger\int \D^3 r \mathlarger \int_0^1 \D s \: \delta\left({\mathbf r} - {\mathbf r}_\alpha\left(s\right) 
   \right) \left(\beta \imun N w\left({\mathbf r}\right) + \beta N U_{\rm s}\left({\mathbf r}\right) \right) \right\} }}
   {\mathlarger \int \mathcal{D} {\mathbf r}_\alpha \left(\bullet \right) \mathcal{P}\left[{\mathbf r}_\alpha 
   \left(\bullet \right) \right]}
   \tag{B.2} \label{eq_appendixB_3}
\end{equation}

From \ref{eq_appendixB_3},
\begin{align}
   \frac{\delta Q}{\delta w}  = & \frac{1}
   {\mathlarger \int \mathcal{D} {\mathbf r}_\alpha \left(\bullet \right) \mathcal{P}\left[{\mathbf r}_\alpha 
   \left(\bullet \right) \right]}
   \mathlarger \int \mathcal{D}{\mathbf r}_\alpha \left(\bullet \right)
   \mathcal{P}\left[{\mathbf r}_\alpha \left(\bullet\right)\right]
   \left[-\beta \imun N \mathlarger \int_0^1 \D s \: \delta\left({\mathbf r}-{\mathbf r}_\alpha \left(s\right) \right) \right]
   \times \nonumber \\
   & \exp{\left[ -\mathlarger\int \D^3 r \mathlarger\int_0^1 \D s \: \delta\left({\mathbf r}-{\mathbf r}_\alpha\left(s\right)\right)
   \left(\beta \imun N w\left({\mathbf r}\right) + \beta N U_{\rm s}\left({\mathbf r}\right) \right) \right]}
   \nonumber \\
   = &  -  \beta\imun N Q
   \left \{ \mathlarger \int \mathcal{D} {\mathbf r}_\alpha \left(\bullet \right) \mathcal{P}\left[{\mathbf r}_\alpha 
   \left(\bullet \right) \right]
   \left[\mathlarger \int_0^1 \D s \: \delta\left({\mathbf r}-{\mathbf r}_\alpha \left(s\right) \right) \right]
   \times \right . \nonumber \\
   & \left . \exp{\left[ -\mathlarger\int \D^3 r \mathlarger\int_0^1 \D s \: \delta\left({\mathbf r}-{\mathbf r}_\alpha\left(s\right)\right)
   \left(\beta \imun N w\left({\mathbf r}\right) + \beta N U_{\rm s}\left({\mathbf r}\right) \right) \right]} \right \}   
   \nonumber \\
   & \mathlarger{\mathlarger /} \left\{ \mathlarger \int \mathcal{D} {\mathbf r}_\alpha \left(\bullet \right) \mathcal{P}\left[{\mathbf r}_\alpha 
   \left(\bullet \right) \right]
   \exp{\left[ -\mathlarger\int \D^3 r \mathlarger\int_0^1 \D s \: \delta\left({\mathbf r}-{\mathbf r}_\alpha\left(s\right)\right)
   \left(\beta \imun N w\left({\mathbf r}\right) + \beta N U_{\rm s}\left({\mathbf r}\right) \right) \right]} \right \}
   \tag{B.3} \label{eq_appendixB_4}
\end{align}
Recalling the definition of $\hat{\phi}$ from  \ref{eq_phihat_definition},  \ref{eq_appendixB_4} can be written 
as: 
\begin{equation}
   \frac{\delta Q}{\delta w} = - \beta N \imun Q \frac{\rho_0}{Nn}
   \left\langle \hat{\phi}\left({\mathbf r}\right)\right\rangle
   \tag{B.4} \label{eq_appendixB_5}
\end{equation}
with $\langle \; \rangle$ denoting an average over the distribution defined by the single-chain partition function
$Q\left[\imun w + U_{\rm s}\right]$,  \ref{eq_appendixB_3}. In averaging $\hat{\phi}\left({\mathbf r}\right)$ of 
\ref{eq_phihat_definition}, all $n$ chains involved in the summation for $\alpha =1$ to $n$ yield the same 
contribution.

We introduce the symbol
\begin{equation}
   \phi\left({\mathbf r}\right) \equiv \left\langle \hat{\phi}\left({\mathbf r}\right) \right\rangle_{n = \bar{n}}
   = \frac{N\bar{n}}{\rho_0} \left\langle \mathlarger\int_0^1 \D s \: 
   \delta\left({\mathbf r} - {\mathbf r}_\alpha \left(s\right) \right) \right\rangle
   \tag{B.5} \label{eq_appendixB_6}
\end{equation}
with $\bar{n}$ being the total number of chains present in the considered interfacial region at the optimum.
Then,  \ref{eq_appendixB_4} gives:
\begin{equation}
   \frac{\delta Q}{\delta w} = - \beta \frac{\rho_0}{\bar{n}} \imun Q \phi\left({\mathbf r}\right)
   \tag{B.6} \label{eq_appendixB_7}
\end{equation}
Substituting  \ref{eq_appendixB_7} in  \ref{eq_appendixB_1}, we obtain:
\begin{equation}
   \frac{\delta H}{\delta w} = -\imun \rho_0 + \kappa_{\rm T} \rho_0^2 w\left({\mathbf r}\right) + 
   \imun \frac{\rho_0}{\bar{n}} \exp{\left(\frac{\mu N}{k_{\rm B}T} \right)} \tilde{N} Z_{\rm free}
   Q\left[ \imun w + U_{\rm s}\right] \phi\left({\mathbf r}\right) = 0
   \tag{B.7} \label{eq_appendixB_8}
\end{equation}
whence the self-consistent field is obtained as:
\begin{equation}
   w \left({\mathbf r}\right) = \frac{\imun}{\kappa_{\rm T} \rho_0} \left\{
   1 - \frac{1}{\bar{n}}\exp{\left(\frac{\mu N}{k_{\rm B}T} \right)} \tilde{N} Z_{\rm free} 
   Q\left[\imun w + U_{\rm s}\right] \phi\left({\mathbf r}\right)
   \right\}
   \tag{B.8} \label{eq_appendixB_9}
\end{equation}

Very far from any interfaces, bulk conditions will prevail:
\begin{align}
   w\left({\mathbf r}\right) & = 0 \nonumber \\
   \phi\left({\mathbf r}\right) & = 1 
   \tag{B.9} \label{eq_appendixB_10}
\end{align}
By applying  \ref{eq_appendixB_9} in that region, we are led to the condition:
\begin{equation}
   \frac{1}{\bar{n}} \exp{\left(\frac{\mu N}{k_{\rm B}T} \right)} \tilde{N} Z_{\rm free}
   Q\left[ \imun w + U_{\rm s}\right] = 1
   \tag{B.10} \label{eq_appendixB_11}
\end{equation}
Using  \ref{eq_appendixB_11} in  \ref{eq_appendixB_9}, we obtain the self-consistent field as
\begin{equation}
   w\left({\mathbf r}\right) = \frac{\imun}{\kappa_{\rm T}\rho_0} \left[ 1 - \phi\left({\mathbf r}\right)\right]
   \tag{B.11} \label{eq_appendixB_12}
\end{equation}

At the optimum,
\begin{equation}
   \mathlarger \int \D^3 r \rho_0 w\left({\mathbf r}\right) = 
   \frac{\imun}{\kappa_{\rm T}} \mathlarger \int \D^3 r \left[1-\phi\left({\mathbf r}\right)\right] =
   \frac{\imun}{\kappa_{\rm T}}\left(V-\frac{\bar{n}N}{\rho_0}\right) = 
   \frac{\imun}{\kappa_{\rm T}} \left(n_{\rm bulk} - \bar{n}\right) \frac{N}{\rho_0}
   \tag{B.12} \label{eq_appendixB_13}
\end{equation}
with $n_{\rm bulk}$ the number of chains within volume $V$ of the bulk polymer under the considered conditions of
$T$ and $\mu N$.

From  \ref{eq_appendixB_12},
\begin{equation}
   \mathlarger \int \D^3 r \frac{\kappa_{\rm T}}{2}\left[ \rho_0 w\left({\mathbf r}\right)\right]^2
   = - \frac{1}{2\kappa_{\rm T}} \mathlarger \int \D^3 r \left[\phi\left({\mathbf r}\right) -1 \right]^2
   \tag{B.13} \label{eq_appendixB_14}
\end{equation}
In view of eqs \ref{eq_appendixB_13} and \ref{eq_appendixB_14}, the first sum in  \ref{eq_eff_hamiltonian} 
of the main text for the field-dependent effective Hamiltonian at the optimum becomes:
\begin{equation}
   \mathlarger \int \D^3 r \left[- \imun \rho_0 w\left({\mathbf r}\right) + \frac{\kappa_{\rm T}}{2} 
   \left(\rho_0 w\left({\mathbf r}\right)\right)^2 \right]
   = \frac{N}{\rho_0 \kappa_{\rm T}} \left(n_{\rm bulk} - \bar{n}\right) 
   - \frac{1}{2 \kappa_{\rm T}} \mathlarger \int \D^3 r\left[\phi\left({\mathbf r}\right) - 1\right]^2
   \tag{B.14} \label{eq_appendixB_15}
\end{equation}
Following a different route, using eqs \ref{eq_appendixB_12} and \ref{eq_appendixB_14}, and the fact that 
$\rho_0 = n_{\rm bulk} N / V$ we obtain an alternative expression for the same quantity as follows:
\begin{align}
   \mathlarger \int \D^3 r \left[-\imun \rho_0 w \left({\mathbf r}\right) + \frac{\kappa_{\rm T}}{2} 
   \left( \rho_0 w \left({\mathbf r}\right)\right)^2\right] = & -\frac{n_{\rm bulk}N}{V}
   \mathlarger \int \D^3 r \: \imun  w \left({\mathbf r}\right) \phi\left({\mathbf r}\right)
   \nonumber \\ 
   & + \frac{1}{2\kappa_{\rm T}} \mathlarger \int \D^3 r \left[1-\phi\left({\mathbf r}\right)\right]^2
   \tag{B.15} \label{eq_appendixB_16}
\end{align}
By virtue of  \ref{eq_appendixB_16}, the effective Hamiltonian, 
\ref{eq_eff_hamiltonian} can be written as:
\begin{align}
   \bar{H} = & 
   - \frac{n_{\rm bulk} N}{V} \mathlarger \int \D^3 r \: \imun
   w\left({\mathbf r}\right) \phi \left({\mathbf r} \right) 
           + \frac{1}{2\kappa_{\rm T}} \mathlarger \int \D^3 r \left[1 - \phi \left({\mathbf r} \right) \right]^2
   \nonumber \\
    & - \frac{1}{\beta} \exp{\left(\frac{\mu N}{k_{\rm B}T} \right)} \tilde{N} Z_{\rm free} 
    Q \left[\imun w + U_{\rm s}\right]
    \tag{B.16} \label{eq_appendixB_17}
\end{align}
where, from  \ref{eq_appendixB_11},
\begin{equation}
   \frac{1}{\beta}\exp{\left(\frac{\mu N}{k_{\rm B}T}\right)} \tilde{N} Z_{\rm free} Q\left[\imun w + U_{\rm s}\right]
   = \frac{\bar{n}}{\beta}
   \tag{B.17} \label{eq_appendixB_18}
\end{equation}
And by setting 
\begin{equation}
   w^\prime\left({\mathbf r}\right) = \imun w \left({\mathbf r}\right) + U_{\rm s} \left({\mathbf r}\right)
   \tag{B18} \label{eq_appendixB_19}
\end{equation}
which is a real field, we arrive at the effective Hamiltonian as expressed in  
\ref{eq_eff_Hamiltonian_optimum} of the main text.

\twocolumn

\begin{acknowledgement} 
Part of this work was funded by the European Union through the project COMPNANOCOMP
under grant number 295355. In addition, we are grateful to the Dutch Polymer Institute (DPI) for financial 
support under project \# 744 (NetSim). 
G.G.V. thanks the Alexander S. Onassis Public Benefit Foundation for a doctoral scholarship.
We thank Dr. Kostas Daoulas for making his SCF code available to us and Ms. Chrysa Charitoglou for some calculations.
\end{acknowledgement} 

\begin{suppinfo}
Discretization scheme and numerical solution of the SCF Edwards diffusion equation.
\end{suppinfo}

\bibliography{scf_tst}

\end{document}